\documentclass[11pt,leqno]{article}

\usepackage[english]{babel}
\usepackage[utf8x]{inputenc}
\usepackage{amsmath,amssymb,amstext,amsthm}
\usepackage{graphicx}
\usepackage[affil-it]{authblk}
\usepackage[colorinlistoftodos]{todonotes}
\usepackage[linktocpage]{hyperref}

\newtheorem{theorem}{Theorem}[section]
\newtheorem{lemma}[theorem]{Lemma}
\newtheorem{corollary}[theorem]{Corollary}
\newtheorem{definition}[theorem]{Definition}

\newtheorem{remark}[theorem]{Remark}

\newcommand{\qeds}{\qed\vspace{.2cm}}

\newcommand{\Q}{\mathcal{Q}}

\newcommand{\A}{\ensuremath{\mathcal{A}}}

\newcommand{\FF}{\ensuremath{\mathcal{F}}}

\DeclareMathOperator{\tr}{Tr}

\DeclareMathOperator{\rel}{rel}

\DeclareMathOperator{\aut}{Aut}
\DeclareMathOperator{\qut}{Qut}

\def\be{\begin{equation}}
\def\ee{\end{equation}}
\usepackage{enumerate}

\title{Nonlocal Games and Quantum Permutation Groups}

\author[1]{Martino Lupini}
\author[2]{Laura Man\v{c}inska} 
\author[3]{David E.~Roberson} 
\affil[1]{Mathematics Department, California Institute of Technology, 1200 East California Boulevard, Mail Code 253-37, Pasadena, CA 91125}
\affil[2]{QMATH, Department of Mathematical Sciences, University of Copenhagen, Universitetsparken 5, 2100 Copenhagen \O, Denmark}
\affil[3]{Department of Applied Mathematics and Computer Science, Technical University of Denmark, DK-2800 Lyngby, Denmark}

\begin{document}

\maketitle

\begin{abstract}
We present a strong connection between quantum information and the theory of quantum permutation groups. Specifically, we define a notion of quantum isomorphisms of graphs based on quantum automorphisms from the theory of quantum groups, and then show that this is equivalent to the previously defined notion of quantum isomorphism corresponding to perfect quantum strategies to the isomorphism game. Moreover, we show that two connected graphs $X$ and $Y$ are quantum isomorphic if and only if there exists $x \in V(X)$ and $y \in V(Y)$ that are in the same orbit of the quantum automorphism group of the disjoint union of $X$ and $Y$.
This connection links quantum groups to the more concrete notion of nonlocal games and physically observable quantum behaviours. In this work, we exploit this by using ideas and results from quantum information in order to prove new results about quantum automorphism groups of graphs, and about quantum permutation groups more generally. In particular, we show that asymptotically almost surely all graphs have trivial quantum automorphism group. Furthermore, we use examples of quantum isomorphic graphs from previous work to construct an infinite family of graphs which are quantum vertex transitive but fail to be vertex transitive, answering a question from the quantum permutation group literature.

Our main tool for proving these results is the introduction of orbits and orbitals (orbits on ordered pairs) of quantum permutation groups. We show that the orbitals of a quantum permutation group form a coherent configuration/algebra, a notion from the field of algebraic graph theory. We then prove that the elements of this \emph{quantum orbital algebra} are exactly the matrices that commute with the magic unitary defining the quantum group. We furthermore show that quantum isomorphic graphs admit an isomorphism of their quantum orbital algebras which maps the adjacency matrix of one graph to that of the other. 

We hope that this work will encourage new collaborations among the communities of quantum information, quantum groups, and algebraic graph theory.

\end{abstract}

\section{Introduction}

An isomorphism of graphs $X$ and $Y$ is function $\varphi : V(X) \to V(Y)$ which is bijective, preserves adjacency, and preserves non-adjacency. Whenever such a function exists we say that $X$ and $Y$ are \emph{isomorphic} and write $X \cong Y$. An alternative definition of isomorphism can be stated in terms of the adjacency matrices of $X$ and $Y$. These are 01-matrices with a $1$ in the entries corresponding to the edges of the graphs. If $X$ and $Y$ have adjacency matrices $A$ and $B$ respectively, then they are isomorphic if and only if there exists a permutation matrix $P$ such that $P^TAP = B$, or equivalently $AP = PB$.

The isomorphisms from a graph $X$ to itself are called \emph{automorphisms} and they form a group called the \emph{automorphism group of $X,$} denoted $\aut(X)$. As with isomorphisms, the automorphisms of a graph can be represented as permutation matrices. Furthermore, the automorphisms of $X$ are exactly those permutations whose corresponding permutation matrix commutes with the adjacency matrix of $X$.

In~\cite{banicahomogeneous}, Banica introduced the \emph{quantum automorphism group} of a graph, generalizing Wang's definition of the \emph{quantum permutation group of a set}~\cite{wang}, which itself used the formalism of Woronowicz's compact quantum groups~\cite{woronowicz}. Briefly, the idea is to consider the algebra of complex-valued functions on the automorphism group of a graph. This algebra is commutative and can be generated by finitely many elements satisfying certain relations depending on the graph. By dropping the (explicit) commutativity requirement of these generators, one obtains a possibly different algebra which is by definition the ``algebra of continuous functions on the \emph{quantum} automorphism group of the graph". The generators can be arranged in a matrix which is called a \emph{magic unitary}, which has the property that it commutes with the adjacency matrix of the graph.

More recently, a different approach was used to define quantum analogs of graph isomorphisms which involved nonlocal games. A (2-party) nonlocal game is a game played by two players, Alice and Bob, against a referee/verifier. The referee sends each of the players some question/input and they must respond with some output. Whether the players win is determined by evaluating a binary verification function that depends on the inputs and outputs of both players. The players have full knowledge of the game beforehand: they know their input and output sets and the verification function, as well as the probability distribution used to dispense the inputs. Their goal is to try to win the game with as high probability as possible. In order to do this, they are allowed to agree on whatever strategy they like beforehand, but are not allowed to communicate after receiving the questions.

A \emph{classical strategy} for a nonlocal game is one in which the only resource available to the players is shared randomness. In a \emph{quantum strategy}, the players are allowed to perform local quantum measurements on a shared entangled state. This does not allow them to communicate, but can sometimes increase their chance of winning. If players are able to win a given nonlocal game with probability greater than what is possible classically, then this is evidence that they are doing something genuinely quantum. Thus nonlocal games provide a way of certifying quantum behaviour through the observation of only classical data: the inputs and outputs of the players. Such quantum behaviour is known as \emph{nonlocality}, and in the extreme case, when quantum players can win with probability 1 but classical players cannot, the game is called a \emph{pseudotelepathy game}~\cite{brassard05}.

In~\cite{qiso1}, along with others, the second and third authors introduced a family of nonlocal games called \emph{isomorphism games}, and investigated the classical and quantum strategies that win the game perfectly (with probability 1). They showed that the game can be won perfectly by a classical strategy if and only if the corresponding graphs are isomorphic. This motivated the definition of quantum isomorphic graphs: those for which the game can be won perfectly by a quantum strategy. They characterized quantum isomorphism in terms of an object they referred to as a \emph{projective permutation matrix}. The entries of this matrix correspond to the quantum measurement operators used to win the game. They showed that if the graphs used in the game had adjacency matrices $A$ and $B$ respectively, then they were quantum isomorphic if and only if there exists projective permutation matrix $\mathcal{P}$ such that $A\mathcal{P} = \mathcal{P}B$. This is a quantum analog of the adjacency matrix formulation of classical isomorphism, since replacing $\mathcal{P}$ with a permutation matrix $P$ recovers the classical definition.

It turns out that the notions of magic unitaries and projective permutation matrices are essentially the same. The only difference is that in the theory of quantum groups the entries are allowed to be elements of any unital $C^*$-algebra, whereas in quantum information the entries are required to be elements of a unital $C^*$-algebra that \emph{admits a trace}. Thus, in this work we define a relaxed notion of quantum isomorphism in which the required projective permutation matrix has entries from some unital $C^*$-algebra, which does not necessarily admit a trace. This is the natural extension of quantum automorphisms from quantum group theory to the setting of isomorphisms. Surprisingly, we show that this is equivalent to the original notion: the existence of a perfect quantum strategy for the isomorphism game. Moreover, we show that connected graphs $X$ and $Y$ are quantum isomorphic if and only if there exists $x \in V(X)$ and $y \in V(Y)$ that are in the same orbit of the quantum automorphism group of the disjoint union of $X$ and $Y$, in perfect analogy to the classical case. Thus the quantum information theoretic notion of quantum isomorphisms can be rephrased completely in terms of quantum automorphism groups. 

We remark that the connection between quantum permutation groups and quantum isomorphisms has very recently been independently noted in~\cite{musto_compositional_2017}. However, the perspective in that work is quite different from ours. While they develop a very broad, and elegant, categorical framework for working with quantum permutations and other similar notions, here we are more focused on using results and ideas from quantum information and algebraic graph theory to establish new results in the theory of quantum permutation groups. Moreover, they focus on a slightly different notion of quantum isomorphisms where the entries of projective permutation matrices are required to be finite dimensional, and which is known to be different from the more general case we consider here. 

\subsection{Outline}

Besides proving the equivalence of our new notion of quantum isomorphism with the notion arising from isomorphism games, the main results of our work are as follows: {\bf (1)} The orbitals of a quantum permutation group form a coherent configuration; {\bf (2)} the value of the Haar state of a quantum permutation group on products of pairs of generators is determined by the orbitals; {\bf (3)} if $X$ and $Y$ are quantum isomorphic, then there is an isomorphism of their quantum orbital algebras that maps the adjacency matrix of $X$ to that of $Y$; {\bf (4)} there exist an infinite number of quantum vertex transitive graphs which are not vertex transitive.

In Section~\ref{sec:background}, we provide background material on the isomorphism game, quantum permutation groups, and coherent configurations/algebras, which are the three main ingredients of this work. In Section~\ref{sec:orbits} we define the orbits and orbitals of an arbitrary quantum permutation group. We then show, in Section~\ref{subsec:equivdef}, that these definitions are equivalent to definitions based on the spaces of ``fixed points" of the action of the quantum permutation group, a notion studied in the quantum group literature. Section~\ref{subsec:acc} contains the proof of our first main result: that the orbitals of a quantum permutation group $\mathcal{Q}$ form a coherent configuration. The span of the characteristic matrices of the classes in this configuration form a coherent algebra which we call the orbital algebra of $\mathcal{Q}$. In Section~\ref{subsec:commprops} we show that the elements of this algebra are exactly those matrices which commute with the magic unitary defining $\mathcal{Q}$. The following Section~\ref{subsec:useca} shows how to make this result practical by using the coherent algebra of the graph, which can be computed in polynomial time.

Section~\ref{sec:qiso} is concerned with quantum isomorphisms, and using pairs of non-isomorphic but quantum isomorphic graphs to construct quantum vertex transitive graphs that fail to be vertex transitive. It begins with Section~\ref{subsec:isoca} in which we prove our second main result: that any quantum isomorphism of two graphs induces an isomorphism of their quantum orbital algebras which maps the adjacency matrix of one to that of the other. In Sections~\ref{subsec:LBCS} and~\ref{subsec:arkhipov} we introduce two previously known constructions whose concatenation, when applied to any connected non-planar graph $Z$, produces a pair of non-isomorphic graphs which are nevertheless quantum isomorphic. In Section~\ref{subsec:vtxtrans} we show that if the graph $Z$ is vertex transitive, then at least one of the graphs produced by this construction is vertex transitive. Finally, we put this all together in Section~\ref{subsec:disjoint} to show how to construct an infinite family of quantum vertex transitive graphs that are not vertex transitive. We end in Section~\ref{sec:discuss} by discussing some possible future directions and interesting open problems.


The idea to use coherent configurations/algebras, a topic from algebraic graph theory, to study quantum automorphism groups is inspired by concurrent work of the second and third authors and others~\cite{qiso2}. They consider a semidefinite relaxation of quantum isomorphisms that they call doubly nonnegative isomorphism. They show that any doubly nonnegative isomorphism of graphs $X$ and $Y$ induces an isomorphism of the coherent algebras of $X$ and $Y$ that maps the adjacency matrix of $X$ to that of $Y$. Such an isomorphism of coherent algebras is called an \emph{equivalence} of $X$ and $Y$, a previously studied notion~\cite{symmpowers}. Conversely, they show that any equivalence of $X$ and $Y$ can be used to construct a doubly nonnegative isomorphism of $X$ and $Y$. Therefore, these two relations are equivalent. This result is what motivated the consideration of coherent algebras for this work.

\section{Background}\label{sec:background}

We begin by giving some background about each of the three main ingredients of this work: the graph isomorphism game, quantum permutation groups, and coherent configurations/algebras.

\subsection{The isomorphism game}\label{subsec:isogame}

In~\cite{qiso1}, a nonlocal game is introduced which captures the notion of graph isomorphism. In turn, by allowing entangled strategies, this allows one to define a type of quantum isomorphism in a natural way. We will give a brief description of this game and its classical/quantum strategies, but for a more thorough explanation we refer the reader to~\cite{qiso1}.

Given graphs $X$ and $Y$, the $(X,Y)$-isomorphism game is played as follows: a referee/verifier sends each of two players (Alice and Bob) a vertex of $X$ or $Y$ (not necessarily the same vertex to both), and tells them which graph it is from (this is necessary if the vertex sets are not disjoint). Each of Alice and Bob must respond to the referee with a vertex of $X$ or $Y$ (and specify which graph it belongs to). Alice and Bob win if they meet two conditions, the first of which is:

\begin{enumerate}[{\bf(1)}]
\item If a player (Alice or Bob) receives a vertex from $X$, they must respond with a vertex from $Y$ and vice versa.
\end{enumerate}

Assuming this condition is met, Alice either receives or responds with a vertex of $X$, which we will call $x_A$, and either responds with or receives a vertex of $Y$, which we will call $y_A$. We can similarly define $x_B$ and $y_B$ for Bob. The second condition they must meet in order to win is then given by

\begin{enumerate}[{\bf(2)}]
\item $\rel(x_A, x_B) = \rel(y_A, y_B)$,
where $\rel$ is a function denoting the \emph{relationship} of two vertices, i.e., whether they are equal, adjacent, or distinct non-adjacent.
\end{enumerate}

The players know the graphs $X$ and $Y$ beforehand and can agree on any strategy they like, but they are not allowed to communicate during the game. For simplicity we may assume that the referee sends the vertices to the players uniformly at random. We only require that Alice and Bob play one round of the game (each receive and respond with a single vertex), but we require that their strategy guarantees that they win with probability 1. We say that such a strategy is a \emph{perfect} or \emph{winning} strategy.

If $\varphi: V(X) \to V(Y)$ is an isomorphism, then it is not difficult to see that responding with $\varphi(x)$ for $x \in V(X)$ and $\varphi^{-1}(y)$ for $y \in V(Y)$ is a perfect strategy for the $(X,Y)$-isomorphism game. Conversely, any perfect deterministic classical strategy can be shown to have this form. This is what was shown in~\cite{qiso1} and it follows that there exists a perfect classical strategy for the $(X,Y)$-isomorphism game if and only if $X \cong Y$. In general, classical players could use shared randomness but it is not hard to see that this would not allow them to win if they were not already able to succeed perfectly using a deterministic strategy.

In a quantum strategy, Alice and Bob have access to a shared entangled state which they are allowed to perform local quantum measurements on. This does not allow them to communicate, but may allow them to correlate their actions/responses in ways not possible for classical players. In~\cite{qiso1}, two different models for performing joint measurements on a shared state were considered: the tensor product framework and the commuting operator framework. In the tensor product framework, both Alice and Bob have their own Hilbert space on which they are allowed to make measurements, and their shared state lives in the tensor product of their Hilbert spaces. We note that for the isomorphism game, whenever there is a perfect quantum strategy in the tensor product framework, there is one that is finite dimensional (this is not known to hold for general nonlocal games). In the commuting operator framework, they share a single Hilbert space in which their shared state lives, but it is required that all of Alice's measurement operators commute with all of Bob's. In either case, the shared entangled state is specified by a vector and the measurement operators are positive operators on the relevant Hilbert spaces. We remark that the commuting operator framework is the more general of the two. This follows from the fact that any tensor product strategy can be made into a commuting operator strategy by replacing each measurement operator $E$ of Alice's with $E \otimes I$, and each measurement operator $F$ of Bob's with $I \otimes F$. However, if we restrict both frameworks to finite dimensional Hilbert spaces then they are known to be equivalent.

In~\cite{qiso1}, they say that graphs $X$ and $Y$ are ``quantum isomorphic" if there exists a perfect quantum strategy for the $(X,Y)$-isomorphism game in the tensor product framework. Similarly, they say they are ``quantum commuting isomorphic" if there exists a perfect quantum strategy in the commuting operator framework. They show that these relations are distinct, and the former is finer than the latter, since the commuting framework is more general. In this work we additionally consider an even more general notion of quantum isomorphism, that we simply call \emph{quantum isomorphism}, which corresponds to the isomorphism game having a perfect $C^*$-strategy as defined in~\cite{vern}. This definition of quantum isomorphism extends the notion of quantum graph automorphisms of Banica. Because of this, we will refer to the relation of being ``quantum isomorphic" from~\cite{qiso1} as being \emph{quantum tensor isomorphic}, and we will use the notation `$\cong_{qt}$' for this relation. We will keep the terminology of quantum commuting isomorphism the same, and use `$\cong_{qc}$' for this.

Without going into the details of the proof, we present the following characterization of quantum tensor isomorphism proven in~\cite{qiso1}. Recall that an operator $P$ is an (orthogonal) \emph{projection} if $P = P^* = P^2$.

\begin{theorem}\label{thm:qtensor}
Let $X$ and $Y$ be graphs. Then $X \cong_{qt} Y$ if and only if there exist projections $P_{xy}$ for $x \in V(X), y \in V(Y)$ on a finite dimensional Hilbert space satisfying
\begin{enumerate}
\item $\sum_{y \in V(Y)} P_{xy} = I$ for all $x \in V(X)$;
\item $\sum_{x \in V(X)} P_{xy} = I$ for all $y \in V(Y)$;
\item $P_{xy}P_{x'y'} = 0$ if $\rel(x,x') \ne \rel(y,y')$.
\end{enumerate}
\end{theorem}

The projections $P_{xy}$ in the above theorem correspond to the measurement operators used by Alice (Bob uses the transpose of these) in a perfect strategy for the isomorphism game. The element $P_{xy}$ corresponds to Alice replying with $y$ upon receiving $x$. Intuitively, this is analogous to $x$ being mapped to $y$ by an isomorphism of $X$ and $Y$.

It is possible to ``compose" strategies for isomorphism games. If $P_{xy}$ are projectors as above for the $(X,Y)$-isomorphism game, and $Q_{yz}$ are projectors for the $(Y,Z)$-isomorphism game, then the construction $R_{xz} := \sum_{y \in V(Y)} P_{xy} \otimes Q_{yz}$ gives projections satisfying Conditions (1)--(3) for the $(X,Z)$-isomorphism game (note the similarity to the definition of comultiplication in a quantum permutation group given in Section~\ref{subsec:qperms}). In terms of the game, this construction corresponds to playing as follows: upon receiving $x \in V(X)$, the player acts as if they were playing the $(X,Y)$-isomorphism game and obtains some outcome $y \in V(Y)$. The player then acts as if $y$ was their input for the $(Y,Z)$-isomorphism game and obtains some output $z \in V(Z)$ which they send to the verifier. Thinking in terms of the gameplay like this can sometimes provide one with good intuition for quantum isomorphisms.

An analogous characterization for quantum commuting isomorphism was also given in~\cite{qiso1}:

\begin{theorem}\label{thm:qcommute0}
Let $X$ and $Y$ be graphs. Then $X \cong_{qc} Y$ if and only if there exists a unital C*-algebra $\mathcal{A}$ that admits a faithful tracial state, and projections $P_{xy} \in \mathcal{A}$ for $x \in V(X), y \in V(Y)$ which satisfy Conditions (1)--(3) of Theorem~\ref{thm:qtensor}.
\end{theorem}


In Theorem \ref{thm:qcommute0} and in the following, we assume a unital C*-algebra to be nontrivial, i.e., such that its unit $1$ is different from the zero element. A \emph{state} on a unital C*-algebra $\mathcal{A}$ is a linear functional $s: \A \to \mathbb{C}$ such that $s(1) = 1$ and $s(A^*A) \ge 0$ for all $A \in \A$. The state is \emph{tracial} if $s(AB) = s(BA)$ for all $A,B \in \A$. The prototypical example of a tracial state is the unique tracial state on the algebra $M_n(\mathbb{C})$ of $n \times n$ complex matrices given by $s(A) = \tr(A)/n$.

In~\cite{qiso1}, it was noted that the projections from Theorem~\ref{thm:qtensor} could be used as the entries of a matrix $U = (P_{xy})\in M_n(\mathcal{A})$ that will satisfy $A U = U B$ where $A$ and $B$ are the adjacency matrices of $X$ and $Y$ respectively. Here and in the following, we canonically regard $M_n(\mathcal{A})$ for a unital C*-algebra $\mathcal{A}$ as an $M_n(\mathbb{C})$-bimodule. In other words, if $A=(a_{ij})\in M_n(\mathcal{A})$ and $B=(b_{ij})\in M_n(\mathbb{C})$, then $AB\in M_n(\mathcal{A})$ is the matrix whose $(i,j)$-entry is $\sum _{i=1}^{n}a_{ik}b_{kj}$, and $BA\in M_n(\mathcal{A})$ is the matrix whose $(i,j)$-entry is $\sum _{i=1}^{n}b_{ik}a_{kj}$.


It was also shown in~\cite{qiso1} that the matrix $U$ described above will necessarily be unitary. This motivated the definition of what they called a ``projective permutation matrix", also known as \emph{magic unitary} in the quantum group literature:

\begin{definition}
A matrix $U = (u_{ij})_{i,j \in [n]}$ whose entries $u_{ij}$ are elements of some unital C*-algebra is a \emph{magic unitary} if $u_{ij}$ is a projection for all $i,j \in [n]$ and $\sum_{j'} u_{ij'} = 1 = \sum_{i'} u_{i'j}$ for all $i,j \in [n]$. Any such matrix will be unitary, i.e., $UU^* = 1 = U^*U$.
\end{definition}

Note that in the case where the C*-algebra is $\mathbb{C}$, a magic unitary is simply a permutation matrix. The notion of projective permutation matrices/magic unitaries allows for the following formulation of quantum tensor isomorphism that was given in~\cite{qiso1}:

\begin{theorem}\label{thm:qtiso}
Let $X$ and $Y$ be graphs with adjacency matrices $A$ and $B$ respectively. Then $X \cong_{qt} Y$ if and only if there exists a magic unitary $U = (u_{xy})_{x \in V(X), y \in V(Y)}$ whose entries are operators on a finite-dimensional Hilbert space such that $AU = UB$.
\end{theorem}

Since any C*-algebra admits a representation as an algebra of operators on a Hilbert space, one can equivalently state Theorem \ref{thm:qtiso} by saying that two graphs are quantum tensor isomorphic if and only if there exists a unital C*-algebra $\mathcal{A}$ that has a finite dimensional representation and magic unitary $U = (u_{xy})_{x \in V(X), y \in V(Y)}$ with entries from $\mathcal{A}$ such that $AU = UB$. Though it was not explicitly written in~\cite{qiso1}, the same reasoning allows for an analogous formulation for quantum commuting isomorphism:

\begin{theorem}\label{thm:qciso}
Let $X$ and $Y$ be graphs with adjacency matrices $A$ and $B$ respectively. Then $X \cong_{qc} Y$ if and only if there exists a magic unitary $U = (u_{xy})$ whose entries are elements of a unital C*-algebra that admits a faithful tracial state such that $AU = UB$.
\end{theorem}


The two theorems above are quantum analogs of the fact that graphs $X$ and $Y$ are isomorphic if and only if there exists a permutation matrix $P$ such that $AP = PB$.

\begin{remark}\label{rem:faithful}
The faithfulness of the tracial state in Theorems~\ref{thm:qcommute0} and~\ref{thm:qciso} is not really necessary. This is because given a (possibly not faithful) tracial state $\tau $ on a $C^*$-algebra $\mathcal{A}$, one can always replace $\mathcal{A}$ with its image under the GNS representation associated with $\tau $, which is endowed with a canonical faithful (vector) state.
\end{remark}

\subsection{Quantum permutation groups}\label{subsec:qperms}

Here we provide some background on quantum groups. Though we aim to give a
brief but thorough summary of quantum groups, the reader is not expected
have have a mastery of these concepts in order to understand this work. The
essential points are the definitions of the quantum symmetric group, the
quantum automorphism group of a graph, and quantum permutation groups in
general, along with their comultiplications, antipodes, counits, and
universal actions.

We adopt the following notation. For subsets $X,Y$ of a C*-algebra $\mathcal{%
A}$, we let $XY$ be the set $\left\{ xy:x\in X,y\in Y\right\} $, and $\left[
X\right] $ be the closure of the linear span of $X$ inside $\mathcal{A}$. We
denote by $\mathcal{A}\otimes \mathcal{B}$ the \emph{minimal} tensor product
of C*-algebras $\mathcal{A}$ and $\mathcal{B}$. Concretely, if $\mathcal{A},%
\mathcal{B}$ are realized as algebra of operaotors on a Hilbert space $%
\mathcal{H}$, then $\mathcal{A}\otimes \mathcal{B}$ can be identified with
the algebra of operators on $\mathcal{H}\otimes \mathcal{H}$ generated by
the elementary tensors of the form $a\otimes b$ for $a\in \mathcal{A}$ and $%
b\in \mathcal{B}$.

A \emph{compact quantum group }$\mathbb{G}$ is given by a unital C*-algebra $%
C(\mathbb{G})$ together with a unital *-homomorphism $\Delta :C(\mathbb{G}%
)\rightarrow C(\mathbb{G})\otimes C(\mathbb{G})$ called \emph{%
comultiplication} satisfying the following identities:

\begin{itemize}
\item $\left( \Delta \otimes \mathrm{id}\right) \circ \Delta =\left( \mathrm{%
id}\otimes \Delta \right) \circ \Delta $;

\item $\left[ \Delta \left( C(\mathbb{G} )\right) \left( 1\otimes C(\mathbb{G%
} )\right) \right] =\left[ \Delta \left( C(\mathbb{G} )\right) \left( C(%
\mathbb{G} )\otimes 1\right) \right] =C(\mathbb{G} )\otimes C\left( \mathbb{G%
}\right) $.
\end{itemize}

Any (classical) compact group $G$ gives rise canonically to a compact
quantum group, where $C\left( G\right) $ is the C*-algebra of continuous
complex-valued functions on $G$, and the comultiplication $\Delta $ is the
function induced by the multiplication operation by duality. In formulas,
one has that, under the identification of $C(G)\otimes C(G)$ with $C(G\times
G)$, $\Delta (f)$ is the function $G\times G\rightarrow \mathbb{C},
(s,t)\mapsto f(st)$. The compact quantum groups $\mathbb{G}$ that arise in
this fashion are precisely those for which the C*-algebra $C(\mathbb{G} )$
is \emph{commutative}. In the general quantum setting, the compact quantum $%
\mathbb{G}$ does not actually exist as a set of point, but it is defined
implicitly by the unital C*-algebra $C(\mathbb{G})$ together with its
comultiplication.

As a further example, one can consider a discrete group $\Gamma $, and let $%
C^{\ast }(\Gamma )$ be the full group C*-algebra. If one denotes by $%
u_{\gamma }$ for $\gamma \in \Gamma $ the canonical generators of $%
C^{\ast}(\Gamma )$ 
one can define a comultiplication on $C^{\ast }\left( \Gamma \right) $ by%
\begin{equation*}
\Delta \left( u_{\gamma }\right) =u_{\gamma }\otimes u_{\gamma }\text{.}
\end{equation*}%
This defines a compact quantum group, which is the \emph{dual }compact
quantum group $\hat{\Gamma}$ of the discrete group $\Gamma $.

One can show that any compact quantum group $\mathbb{G}$ is endowed with a
canonical state $h:C(\mathbb{G})\rightarrow \mathbb{C}$, called the \emph{%
Haar state}, which is \emph{invariant }in the sense that $\left( h\otimes 
\mathrm{id}\right) \circ \Delta =\left( \mathrm{id}\otimes h\right) \circ
\Delta =h$ (where we identify $\mathbb{C}$ with the scalar multiples of the
identity in $C(\mathbb{G})$).\ In the case of classical groups, this
corresponds to integration with respect to the Haar probability measure.



A (unitary) \emph{representation }of dimension $n$ of a compact quantum
group $\mathbb{G}$ is an $n\times n$ matrix $u=\left[ u_{ij}\right] \in
M_{n}\left( C(\mathbb{G})\right) $ such that $\Delta \left( u_{ij}\right)
=\sum_{k=1}^{n}u_{ik}\otimes u_{kj}$. One can also regard $u$ as a linear
map $u:\mathbb{C}^{n}\rightarrow \mathbb{C}^{n}\otimes C\left( \mathbb{G}%
\right) $ defined by $u\left( e_{i}\right) =\sum_{j=1}^{n}e_{j}\otimes u_{ji}
$. A subspace $\mathcal{K}$ of $\mathbb{C}^{n}$ is \emph{invariant }for the
representation $u$ of $\mathbb{G}$ if $u$ maps $\mathcal{K}$ to $\mathcal{K}%
\otimes C(\mathbb{G})$. A finite-dimensional unitary representation of $%
\mathbb{G}$ is called \emph{irreducible }if it has no nontrivial invariant
subspace. Every (finite-dimensional) representation of $\mathbb{G}$ can be
written as a direct sum of irreducible representations. We will denote by 
\textrm{Rep}$(\mathbb{G})$ the set of finite-dimensional unitary
representations of $\mathbb{G}$, and by \textrm{Irr}$(\mathbb{G})\subset 
\mathrm{Rep}(\mathbb{G})$ the set of irreducible representations of $\mathbb{%
G}$. For $\lambda \in \mathrm{Rep}(\mathbb{G})$ we let $d_{\lambda }$ be the
corresponding dimension, and $u^{\lambda }\in M_{d_{\lambda }}(C(\mathbb{G}))
$ be the matrix defining defining the representation. The \emph{matrix
coefficients }of a unitary representation $\lambda $ of $\mathbb{G}$ are the
elements of $C(\mathbb{G})$ of the form $u_{\xi ,\eta }^{\lambda
}=\left\langle \xi ,u^{\lambda }\left( \eta \right) \right\rangle $ for $\xi
,\eta \in \mathbb{C}^{d_{\lambda }}$.

One then lets $\mathcal{O}(\mathbb{G})$ be the subset of $C(\mathbb{G})$
given by%
\begin{equation*}
\left\{ u_{\xi ,\eta }^{\lambda }:\lambda \in \mathrm{Rep}\left( \mathbb{G}%
\right) ,\xi ,\eta \in \mathbb{C}^{d_{\lambda }}\right\} \text{,}
\end{equation*}%
which is a dense self-adjoint subalgebra of $C(\mathbb{G})$ invariant under
the comultiplication. If $e_{1},\ldots ,e_{d_{\lambda }}$ is the canonical
basis of $\mathbb{C}^{d_{\lambda }}$, then the comultiplication is defined
by $\Delta (u_{\xi ,\eta }^{\lambda })=u_{\xi ,e_{1}}^{\lambda }\otimes
u_{e_{1},\eta }^{\lambda }+\cdots +u_{\xi ,e_{d_{\lambda }}}^{\lambda
}\otimes u_{e_{d_{\lambda }},\eta }^{\lambda }$. The *-algebra $\mathcal{O}%
\left( \mathbb{G}\right) $ is furthermore endowed with a canonical Hopf
*-algebra structure, which is determined, together with the
comultiplication, by the \emph{counit }$\varepsilon :\mathcal{O}(\mathbb{G}%
)\rightarrow \mathbb{C}$ and \emph{antipode }maps $S:\mathcal{O}(\mathbb{G}%
)\rightarrow \mathcal{O}(\mathbb{G})^{\mathrm{op}}$ given by $\varepsilon
(u_{\xi ,\eta }^{\lambda })=\left\langle \xi ,\eta \right\rangle $ and $%
S(u_{\xi ,\eta }^{\lambda })=(u_{\eta ,\xi }^{\lambda })^{\ast }$ for $%
\lambda \in \mathrm{Rep}(\mathbb{G})$ and $\xi ,\eta \in \mathcal{H}%
^{\lambda }$. This means that $\varepsilon $ is a *-homomorphism and $S$ is
a conjugate linear unital anti-homomorphism satisfying%
\begin{equation*}
m\circ \left( S\otimes \mathrm{id}\right) \circ \Delta =m\circ \left( 
\mathrm{id}\otimes S\right) \circ \Delta =\varepsilon 
\end{equation*}%
and%
\begin{equation*}
\left( \varepsilon \otimes \mathrm{id}\right) \circ \Delta =\left( \mathrm{id%
}\otimes \varepsilon \right) \circ \Delta =\mathrm{id}\text{,}
\end{equation*}%
where $m:\mathcal{O}\left( \mathbb{G}\right) \otimes \mathcal{O}\left( 
\mathbb{G}\right) \rightarrow \mathcal{O}\left( \mathbb{G}\right) $ is the
linear map induced by mulitplication, and where we identify $\mathbb{C}$
with the scalar multiples of the identity within $C\left( \mathbb{G}\right) $%
.

In the case of $C\left( G\right) $ for a classical group $G$, one has that $%
C\left( G\right) =\mathcal{O}\left( G\right) $, and the counit and antipode
are given by $\varepsilon \left( f\right) =f\left( 1_{G}\right) $ and $%
S\left( f\right) =f\circ \mathrm{inv}$, where $1_{G}$ is the identity
element of $G$ and $\mathrm{inv}$ is the function from $G$ to itself mapping
each element to its inverse. A complete reference for the theory of compact
quantum groups can be found in \cite{timmermann_invitation_2008}.

Fix a finite set $X$. The \emph{quantum symmetric group on $X$}, denoted $%
\mathcal{Q}_{X}$, is a compact quantum group defined by Wang \cite{wang},
and can be regarded as the quantum analog of the group $S_{X}$ of
permutations of $X$. By definition, such a compact quantum group is defined
by letting $C(\mathcal{Q}_{X})$ be the universal C*-algebra generated by
elements $u_{xy}$ for $x,y\in X$, subject to the relation that the matrix $%
U=(u_{xy})$ be a magic unitary, with comultiplication defined by setting $%
\Delta \left( u_{xz}\right) =\sum_{y\in X}u_{xy}\otimes u_{yz}$. In this
case, one has that $C(\mathcal{Q}_{X})$ is the universal C*-algebra
generated by $\mathcal{O}(\mathcal{Q}_{X})$, and the antipode and counit
extends from $\mathcal{O}(\mathcal{Q}_{X})$ to bounded linear maps on $C(%
\mathcal{Q}_{X})$ defined by $S\left( u_{xy}\right) =u_{yx}$ and $%
\varepsilon \left( u_{xy}\right) =\delta _{xy}$. In this case, the antipode
is an involutive *-linear map, which by definition
means that $\mathcal{O}(\mathcal{Q}_{X})$ is a Hopf *-algebra of \emph{Kac
type}. For comparison, one should consider the C*-algebra $C(S_{X})$
associated with the symmetric group $S_{X}$. This can be seen as the
universal C*-algebra generated by elements $u_{xy}$ for $x,y\in X$ subject
to the relations that $U=(u_{xy})$ be a magic unitary with \emph{pairwise
commuting entries}. In this case, the canonical Hopf *-algebra structure is
defined by the same formulas as in the case of $\mathcal{Q}_{X}$. The
canonical generators $u_{xy}$ of $C(S_{X})$ can be identified with the
elements given by the characteristic functions of the set of permutations
mapping $x$ to $y$ for $x,y\in X$. When $X$ is a finite set with $n$
elements, we also denote $S_{X}$ with $S_{n}$ and $\mathcal{Q}_{X}$ with $%
\mathcal{Q}_{n}$.

The quantum symmetric group $\mathcal{Q}_{X}$ admits a natural universal
property. Recall that a \emph{compact} \emph{quantum space }$\mathbb{X}$ is
given by a unital C*-algebra $C\left( \mathbb{X}\right) $. An \emph{action }%
of a compact quantum group $\mathbb{G}$ on $\mathbb{X}$ is given by a
*-homomorphism $\alpha :C\left( \mathbb{X}\right) \rightarrow C\left( 
\mathbb{X}\right) \otimes C(\mathbb{G})$ satisfying:

\begin{itemize}
\item $\left( \alpha \otimes \mathrm{id}\right) \circ \alpha =\left( \mathrm{%
id}\otimes \Delta \right) \circ \alpha $;

\item $\left[ \left( 1\otimes C(\mathbb{G} )\right) \alpha \left( C(\mathbb{X%
})\right) \right] =C\left( \mathbb{X}\right) \otimes C(\mathbb{G} )$.
\end{itemize}

Given such an action of $\mathbb{G}$ on $\mathbb{X}$, the quantum orbit
space $\mathbb{X}/\mathbb{G}$ is by definition the compact quantum space
such that $C(\mathbb{X}/\mathbb{G})$ is the C*-subalgebra $C\left( \mathbb{X}%
\right) ^{\mathbb{G}}=\left\{ a\in C\left( \mathbb{X}\right) :\alpha \left(
a\right) =a\otimes 1\right\} $ of $C\left( \mathbb{X}\right) $ (\emph{fixed
point algebra}). There is a canonical \emph{conditional expectation }%
(nondegenerate $C\left( \mathbb{X}/\mathbb{G}\right) $-bimodular completely
positive map) $\mathrm{E}_{\mathbb{X}/\mathbb{G}}:C\left( \mathbb{X}\right)
\rightarrow C\left( \mathbb{X}/\mathbb{G}\right) $ given by $\mathrm{E}_{%
\mathbb{X}/\mathbb{G}}=\left( \mathrm{id}\otimes h\right) \circ \alpha $.
The action $\alpha $ is called \emph{transitive }or \emph{homogeneous }if $%
C\left( \mathbb{X}/\mathbb{G}\right) =C\left( \mathbb{X}\right) $. More
information concerning actions of compact quantum groups on compact quantum
spaces can be found in \cite{de_commer_actions_2017}.

Any classical compact Hausdorff space $X$ can in particular be regarded as a
compact quantum space, by considering the algebra $C\left( X\right) $ of
continuous complex-valued functions on $X$. If $G$ is a classical compact
group, then any classical action of $G$ on $X$ yields an action as defined
above, and all the actions arise in this fashion. In this case, $C\left(
X/G\right) $ coincides with the algebra of continuous functions on the
quotient space $X/G$ defined in the usual way as the space of $G$-orbits
endowed with the quotient topology. Furthermore, the canonical conditional
expectation $\mathrm{E}_{X/G}:C\left( X\right) \rightarrow C\left(
X/G\right) $ is given by averaging with respect to the Haar probability
measure on $X$.

In the particular case when $X$ is a finite space, one has that $C\left(
X\right) \cong \mathbb{C}^{X}$. The symmetric group $S_{X}$ admits a
canonical action on $X$ which has the following universal property: for
every compact group $G$ and action of $G$ on $X$, there exists a unique
group morphism $G\rightarrow S_{X}$ that intertwines the actions. Similarly,
the quantum symmetric group $\mathcal{Q}_{X}$ admits a canonical action $%
\alpha _{u}$ on $X$, defined by $\alpha _{u}:\mathbb{C}^{X}\rightarrow 
\mathbb{C}^{X}\otimes C\left( \mathcal{Q}_X\right) $, $e_{x}\mapsto
\sum_{y}e_{y}\otimes u_{yx}$, where $\left( e_{x}\right) _{x\in X}$ denotes
the canonical basis of $\mathbb{C}^{X}$. As shown in \cite{wang}, such an
action satisfies an analog universal property within the class of actions of
compact quantum groups on $X$. If $\mathbb{G}$ is a compact quantum group,
and $\alpha $ is an action of $\mathbb{G}$ on $X$, then there exists a
quantum group morphism from $\mathbb{G}$ to $\mathcal{Q}_{X}$ that
intertwines $\alpha $ and $\alpha _{u}$, i.e., a unital *-homomorphism $%
\varphi :C(\mathcal{Q}_{X})\rightarrow C(\mathbb{G}) $ such that $\Delta
\circ \varphi =\left( \varphi \otimes \varphi \right) \circ \Delta $ and $%
\alpha \circ \varphi =\left( \mathrm{id}\otimes \varphi \right) \circ \alpha
_{u}$.

One can generalize such a construction to define the quantum automorphism
groups of graphs. Suppose now that $X$ is a finite graph with vertex set $%
V(X)$ and adjacency matrix $A$. The (classical) automorphism group $\mathrm{%
Aut}\left( X\right) $ of $X$ is the subgroup of $S_{V(X)}$ consisting of
permutations that map pairs of adjacent vertices in $X$ to pairs of adjacent
vertices in $X$. One can see that the algebra $C\left( \mathrm{Aut}\left(
X\right) \right) $ of continuous functions on $\mathrm{Aut}\left( X\right) $
can be regarded as the universal C*-algebra generated by elements $u_{xy}$
for $x,y\in V\left( X\right) $ subject to the relation that $U=(u_{xy})$ is
a magic unitary with pairwise commuting entries such that $AU=UA$. The
latter relation is equivalent to $u_{xy}u_{x^{\prime }y^{\prime }}=0$ for
vertices $x,x^{\prime },y,y^{\prime }$ of $X$ such that $\rel(x,x^{\prime
})\neq \rel(y,y^{\prime })$, as shown in \cite{fulton_quantum_nodate}.

The \emph{quantum automorphism group }of $X$, which we denote by\textrm{\ Qut%
}$\left( X\right) $, is a compact quantum group defined by letting $C\left( 
\mathrm{Qut}\left( X\right) \right) $ be the universal C*-algebra generated
by elements $u_{xy}$ for $x,y\in V(X)$, subject to the relations that $%
U=(u_{xy})$ is a magic unitary such that $AU=UA$, with comultiplication
given by $\Delta \left( u_{xz}\right) =\sum_{y\in V(X)}u_{xy}\otimes u_{yz}$
for $x,z\in V(X)$. Clearly, in the particular case when $X$ is the graph on $%
n$ vertices without edges, $\mathrm{Qut}\left( X\right) $ recovers the
quantum symmetric group $\mathcal{Q}_{V(X)}$. We also denote $C\left( 
\mathrm{Qut}\left( X\right) \right) $ by $\mathcal{A}\left( X\right) $, and
call it the \emph{quantum symmetry algebra }of the graph $X$. Again, in this
case one has that $C\left( \mathrm{Qut}\left( X\right) \right) $ is the
universal C*-algebra generated by $\mathcal{O}\left( \mathrm{Qut}\left(
X\right) \right) $, and the counit and antipode of $\mathcal{O}\left( 
\mathrm{Qut}\left( X\right) \right) $ extends to bounded linear maps on $%
C\left( \mathrm{Qut}\left( X\right) \right) $. Furthermore, in this case the
antipode is a *-linear map, i.e., $\mathrm{Qut}\left( X\right) $ is of Kac
type. In particular, this implies that the Haar state on $\mathrm{Qut}\left( X\right) $ is \emph{tracial}.

The canonical surjective unital *-homomorphism from $C(\mathcal{Q}_{V(X)})$
to $C\left( \mathrm{Qut}\left( X\right) \right) $ commutes with the
comultiplication, and witnesses that $\mathrm{Qut}\left( X\right) $ is a 
\emph{quantum subgroup }of $\mathcal{Q}_{V(X)}$. For similar reasons, the
(classical) automorphism group $\mathrm{Aut}\left( X\right) $ is a quantum
subgroup of $\mathrm{Qut}\left( X\right) $. More generally, given a finite
set $X$, we say that a compact quantum group $\mathbb{G}$ is a \emph{quantum
permutation group acting on} $X$ if it is a closed quantum subgroup of $%
\mathcal{Q}_{X}$, i.e., there exists a surjective unital *-homomorphism $C(%
\mathcal{Q}_{X})\rightarrow C\left( \mathbb{G}\right) $. In this case, we
call the image of the canonical generators of $C(\mathcal{Q}_{X})$ inside $C(%
\mathbb{G})$ the canonical generators of $C(\mathbb{G})$. We remark that
these form a magic unitary $u= (u_{xy})$, which we call the magic unitary
generating $C(\mathbb{G})$. The \emph{universal action }$u$ of $\mathbb{G}$
on $X$ is the action of $\mathbb{G}$ on $X$ given by $u:\mathbb{C}%
^{X}\rightarrow \mathbb{C}^{X}\otimes C(\mathbb{G})$, $e_{x}\mapsto
\sum_{y\in X}e_{y}\otimes u_{yx}$.

One says that a finite graph $X$ \emph{has no quantum symmetry }if $\mathrm{%
Aut}\left( X\right) =\mathrm{Qut}\left( X\right) $, i.e., the canonical
unital *-homomorphism from $C\left( \mathrm{Qut}\left( X\right) \right) $
onto $C\left( \mathrm{\mathrm{Aut}}\left( X\right) \right) $ is injective.
This is equivalent to the assertion that $C\left( \mathrm{Qut}\left(
X\right) \right) $ is a commutative C*-algebra. It is shown in \cite{wang}
that, for $n\in \left\{ {1,2,3}\right\} $ the graph with no edges and $n$
vertices has no quantum symmetry, i.e., $\mathcal{Q}_{n}=S_{n}$. However,
already for $n=4$ one has that the group $\mathcal{Q}_{n}$ is not classical,
and in fact $C(\mathcal{Q}_{n})$ is noncommutative and infinite-dimensional.
For $n\leq 3$, the lack of any quantum symmetry follows from the fact that
the entries of the magic unitary necessarily commute. This is trivial for $%
n=1$, and it remains straighforward for $n=2$ since any $2\times 2$ magic
unitary has the form 
\begin{equation*}
\begin{pmatrix}
p & 1-p \\ 
1-p & p%
\end{pmatrix}%
\end{equation*}%
For $n=3$, most papers claim that the proof is \textquotedblleft tricky" and
makes use of the Fourier transform over $\mathbb{Z}_{3}$ and the universal
action. However, here we present a short elementary proof. Another
elementary, but longer proof, can be found in~\cite{ji12}, where it is shown
that the operators used to ``quantum 3-color'' the endpoints of an edge must
commute.

Suppose that $U=(u_{ij})_{i,j\in [3]}$ is a magic unitary. It suffices to
show that $u_{ij}$ and $u_{\ell k}$ commute for $i\neq \ell $ and $j\neq k$,
because all other pairs must commute (their product in either direction is
zero) by the definition of magic unitaries. Since we can permute the rows
and columns of a magic unitary independently and always produce a magic
unitary, it suffices to show that $u_{11}u_{22}=u_{22}u_{11}$. We have that 
\begin{equation*}
u_{11}u_{22}=u_{11}u_{22}(u_{11}+u_{12}+u_{13})=u_{11}u_{22}u_{11}+u_{11}u_{22}u_{13},
\end{equation*}%
but 
\begin{equation*}
u_{11}u_{22}u_{13}=u_{11}(1-u_{21}-u_{23})u_{13}=u_{11}u_{13}=0.
\end{equation*}%
Therefore $u_{11}u_{22}=u_{11}u_{22}u_{11}$ and thus $u_{22}u_{11} =
u_{11}u_{22}$ by applying $^*$ to this equation. 

\subsection{Coherent configurations/algebras}

\label{subsec:cohalgs}

A \emph{coherent configuration}~\cite{higman} on a set $X$ is a partition $%
\mathcal{R} = \{R_i : i \in \mathcal{I}\}$ of $X \times X$ into the \emph{%
relations} (or \emph{classes}) $R_i$ satisfying the following:

\begin{enumerate}
\item There is a subset $\mathcal{D} \subseteq \mathcal{I}$ of the index set
such that $\{R_d : d \in \mathcal{D}\}$ is a partition of the diagonal $%
\{(x,x) : x \in X\}$. These are called the \emph{diagonal relations}.

\item For each $R_i$, its converse $\{(y,x) : (x,y) \in R_i\}$ is a
relation, say $R_{i^{\prime }}$, in $\mathcal{R}$. Note that $R_{i^{\prime
}}=R_i$ is possible.

\item For all $i,j,k \in \mathcal{I}$ and any $(x,z) \in R_k$, the number of $y
\in X$ such that $(x,y) \in R_i$ and $(y,z) \in R_j$ is a constant $p_{ij}^k$
that does not depend on $x$ and $z$. We refer to the $p_{ij}^k$ as the \emph{%
intersection numbers} of $\mathcal{R}$.
\end{enumerate}

The motivation here is that the orbits on $X \times X$ of any permutation
group acting on $X$ gives a coherent configuration, but the converse does
not necessarily hold.

Given any coherent configuration, one can construct the matrices $A^i$ for $%
i \in \mathcal{I}$ such that 
\begin{equation*}
A^i_{x,y} = 
\begin{cases}
1 & \text{if } (x,y) \in R_i \\ 
0 & \text{otherwise.}%
\end{cases}
\end{equation*}
We will refer to the matrices $A^i$ as the \emph{characteristic matrices} of
the configuration. One can consider the linear span of the matrices $A^i$,
and it turns out that this will be a \emph{coherent algebra}: a
self-adjoint, unital algebra containing the all ones matrix and which is
closed under Schur (entrywise) product. It is easy to see that most of these
properties follow from the definition of a coherent configuration, with the
possible exception of the fact that the span of the $A^i$ is indeed closed
under matrix multiplication. This closure follows from property (3) above
which can be written in terms of the characteristic matrices as $A^i A^j =
\sum_k p_{ij}^k A^k$ for all $i,j \in \mathcal{I}$.

Conversely, it is a short exercise to show that any coherent algebra necessarily has a unique basis of 01-matrices (this follows from being closed under Schur product) which define
a partition of $X \times X$ (since a coherent algebra contains the all ones
matrix). It is then straightforward to check that this partition satisfies
the conditions of being a coherent configuration. Thus, coherent
configurations and coherent algebras are really just two perspectives on the
same idea.

An important case of coherent configurations/algebras is the coherent
algebra of a graph. It is not difficult to see that the intersection of two
coherent algebras is again a coherent algebra. In terms of coherent
configurations, the intersection of the algebras corresponds to the join (in
the sense of partitions) of the two initial coherent configurations. Because
of this, there is a unique smallest coherent algebra containing the
adjacency matrix of a given graph, and this is called the coherent algebra
of the graph. In terms of configurations, this is the coarsest coherent
configuration such that the edge set of the graph is a union of its
relations. We remark that this coherent algebra can be computed in
polynomial time via the (2-dimensional) Weisfeiler-Leman method~\cite{WL}.%

\section{Quantum orbits and orbitals}\label{sec:orbits}

Given a group $\Gamma$ acting on a finite set $X$ via the action $(\gamma ,x)\mapsto \gamma \cdot x$, one can define its orbit equivalence relation $\sim$ on $X$ as $x \sim y$ if there exists $\gamma \in \Gamma$ such that $\gamma \cdot x = y$. It is easy to see that this is an equivalence relation whose equivalence classes are the orbits of $\Gamma$. Moreover, the action of $\Gamma$ on $X$ induces \emph{diagonal action} on $X \times X$ given by $\gamma \cdot (x,y) = (\gamma \cdot x, \gamma \cdot y)$. One can similarly define the orbits of this action, and these are often referred to as the \emph{orbitals} of the $\Gamma$-action on $X$. We aim to define and investigate analogs of orbits and orbitals for the quantum automorphism groups of graphs.

\begin{definition}
Suppose $X$ is a finite set and $\mathcal{Q}$ is a quantum permutation group acting on $X$. Let $U = (u_{xy})_{x,y \in X}$ be a magic unitary defining $C(\mathcal{Q})$. Define the relations $\sim_1$ and $\sim_2$ on $X$ and $X \times X$ respectively as follows:
\begin{itemize}
\item $x \sim_1 y$ if $u_{xy} \ne 0$;
\item $(x,x') \sim_2 (y,y')$ if $u_{xy}u_{x'y'} \ne 0$.
\end{itemize}
\end{definition}

Of course, $x \sim_1 y$ is meant to be analogous to there being some permutation that maps $x$ to $y$, and similarly $(x,x') \sim_2 (y,y')$ is meant to be analogous to there being some permutation that maps $x$ to $y$ while also mapping $x'$ to $y'$. We will see that the analogy is fairly strong.

We will show that $\sim_1$ and $\sim_2$ are equivalence relations, and then define the orbits/orbitals of $\mathcal{Q}$ on $X$ as their equivalence classes. In the course of the proof, we will use the Hopf *-algebra structure of $\mathcal{O}(\mathcal{Q})$ and the fact that the canonical generators $u_{xy}$ belong to $\mathcal{O}(\mathcal{Q})$.

\begin{lemma}
Given a quantum permutation group $\mathcal{Q}$ acting on a finite set $X$, the relation $\sim_1$ defined above is an equivalence relation.
\end{lemma}
\proof
To see that $\sim_1$ is reflexive, recall that the counit, $\varepsilon$, of a quantum permutation group is given by $\varepsilon(u_{xy}) = \delta_{xy}$, and is a $*$-homomorphism. Therefore, $\varepsilon(u_{xx}) = 1$ implies that $u_{xx} \ne 0$. Therefore $x \sim_1 x$.

To see that $\sim_1$ is symmetric, recall that the antipode $S$ of $\mathcal{Q}$ is given by $S(u_{xy}) = u_{yx}$, and it is an involutive anti-homomorphism. Therefore, if $u_{xy} \ne 0$, then $S^2(u_{xy}) = u_{xy} \ne 0$ implies that $u_{yx} = S(u_{xy}) \ne 0$. In other words, $x \sim_1 y$ implies that $y \sim_1 x$.

Now suppose that $u_{xy} \ne 0$ and $u_{yz} \ne 0$. Recall that the comultiplication $\Delta: C(\mathcal{Q}) \to C(\mathcal{Q}) \otimes C(\mathcal{Q})$ of $\mathcal{Q}$ is a $*$-homomorphism defined as $\Delta(u_{xz}) = \sum_w u_{xw} \otimes u_{wz}$. Thus
\begin{align*}
(u_{xy} \otimes u_{yz})\Delta(u_{xz}) &=  (u_{xy} \otimes u_{yz}) \sum_w u_{xw} \otimes u_{wz} \\
&= \sum_w u_{xy}u_{xw} \otimes u_{yz}u_{wz} = u_{xy} \otimes u_{yz} \ne 0.
\end{align*}
Therefore $\Delta(u_{xz}) \ne 0$ and so $u_{xz} \ne 0$. This shows that if $x \sim_1 y$ and $y \sim_1 z$, then $x \sim_1 z$, i.e., the relation $\sim_1$ is transitive.

Altogether, the above implies that $\sim_1$ is an equivalence relation as desired.\qeds

The above lemma allows us to give the following definition:

\begin{definition}
Given a quantum permutation group $\mathcal{Q}$ acting on a finite set $X$, the \emph{orbits} of $\mathcal{Q}$ are the equivalence classes of the relation $\sim_1$ defined above. In the case where $\mathcal{Q}$ is the quantum automorphism group of a graph $X$, we will sometimes refer to its orbits as the \emph{quantum orbits} of $X$, in order to distinguish them from the orbits of its (classical) automorphism group.
\end{definition}

We remark that orbits of quantum permutation groups were independently introduced and studied by Banica and Freslon in the recent paper~\cite{banicaorbits}. To our knowledge, this is the first time that orbitals have been considered in the quantum setting.

Now we will show that $\sim_2$ is an equivalence relation as we did for $\sim_1$.

\begin{lemma}
Given a quantum permutation group $\mathcal{Q}$ acting on a finite set $X$, the relation $\sim_2$ defined above is an equivalence relation. 
\end{lemma}
\proof
Since $\varepsilon(u_{xx}u_{yy}) = \varepsilon(u_{xx})\varepsilon(u_{yy}) = 1\cdot 1 = 1$, we have that $u_{xx}u_{yy} \ne 0$. Therefore $(x,y) \sim_2 (x,y)$, i.e., $\sim_2$ is reflexive.

If $(x,x') \sim_2 (y,y')$, then $u_{xy}u_{x'y'} \ne 0$ and hence $u_{x'y'}u_{xy} = (u_{xy}u_{x'y'})^* \ne 0$. Since the antipode $S$ is an involution, this implies that
\[
0 \ne S(u_{x'y'}u_{xy})=u_{yx}u_{y'x'}
\]
and thus $(y,y') \sim_2 (x,x')$. So we have shown that $\sim_2$ is symmetric.

Now suppose that $(x,x') \sim_2 (y,y')$ and $(y,y') \sim_2 (z,z')$, i.e., that $u_{xy}u_{x'y'} \ne 0 \ne u_{yz}u_{y'z'}$. Then
\begin{align*}
\Delta(u_{xz}u_{x'z'}) &= \Delta(u_{xz})\Delta(u_{x'z'}) = \left(\sum_w u_{xw} \otimes u_{wz}\right)\left(\sum_{w'} u_{x'w'} \otimes u_{w'z'}\right) = \sum_{w,w'} u_{xw}u_{x'w'} \otimes u_{wz}u_{w'z'}.
\end{align*}
Therefore,
\begin{align*}
(u_{xy} \otimes u_{yz}) \Delta(u_{xz}u_{x'z'}) (u_{x'y'} \otimes u_{y'z'}) &= \sum_{w,w'} u_{xy}u_{xw}u_{x'w'}u_{x'y'} \otimes u_{yz}u_{wz}u_{w'z'}u_{y'z'} \\
&= u_{xy}u_{x'y'} \otimes u_{yz}u_{y'z'} \ne 0
\end{align*}
since each factor in the tensor product is nonzero by assumption. This implies that $\Delta(u_{xz}u_{x'z'}) \ne 0$ and therefore $u_{xz}u_{x'z'} \ne 0$, i.e., $(x,x') \sim_2 (z,z')$. This shows that $\sim_2$ is transitive and thus we have shown it is an equivalence relation.\qeds

Now we can consider the following:

\begin{definition}
Given a quantum permutation group $\mathcal{Q}$ acting on a finite set $X$, the \emph{orbitals} of $\mathcal{Q}$ are the equivalence classes of the relation $\sim_2$ defined above. In the case where $\mathcal{Q}$ is the quantum automorphism group of a graph $X$, we will sometimes refer to its orbitals as the \emph{quantum orbitals} of $X$.
\end{definition}

\begin{remark}\label{rem:orbitalsagree}
Using the fact that $u_{xy}^2 = u_{xy}$, we have that the relation $\sim _1$ ``agrees" with $\sim _2$, when $X$ is identified with the diagonal of $X \times X$. Since the diagonal is invariant with respect to $\sim _2$, one can say that the orbits are precisely the orbitals contained in the diagonal.
\end{remark}


\subsection{An equivalent definition}\label{subsec:equivdef}


In this section, we verify that the notion of quantum orbits from the previous subsection coincides with the usual notion of quantum orbits for an action of a compact quantum group on a compact (quantum) space. This has also been shown in \cite{banicaorbits}. Here we additionally provide an analogous characterization of the orbitals. This might be surprising, as in general the ``quantum orbital space'' of a compact quantum group action is not well defined, as there does not exist in general a quantum analog of the ``diagonal action'' associated with a given action.

Let $\mathcal{Q}$ be a quantum permutation group on a set $X$, and let $U = (u_{ij})$ be the magic unitary whose entries are the canonical generators of $C(\mathcal{Q})$. Recall that the \emph{universal action} $u: \mathbb{C}^X \to \mathbb{C}^X \otimes C(\mathcal{Q})$ of $\mathcal{Q}$ is the linear map defined on the standard basis vectors as
\[u(e_i) = \sum_{j \in X} e_j \otimes u_{ji}.\]

The corresponding quantum orbit space is given by the selfadjoint subalgebra $(\mathbb{C}^X)^{\mathcal{Q}}$ of $\mathbb{C}^{X}$ defined as those elements $f \in \mathbb{C}^X$ such that $u(f)=f\otimes 1$. We say that $(\mathbb{C}^X)^{\mathcal{Q}}$ is \emph{trivial} if it only contains the scalar multiples of $1 \in \mathbb{C}^X$ (the vector with all entries equal to one). As $u$ is a $*$-homomorphism, $(\mathbb{C}^X)^{\mathcal{Q}}$ is a C*-subalgebra of $\mathbb{C}^X$, and therefore has a unique basis of 01-vectors which corresponds to a partition of $X$. We will show that the classes of this partition are precisely the orbits of $\mathcal{Q}$ as defined in the previous section. We note that $(\mathbb{C}^X)^{\mathcal{Q}}$ can be also seen as the space $\mathrm{Hom}(1,u)$ of intertwiners from the trivial representation $1$ to $u$.

\begin{lemma}\label{lem:orbits2orbits}
Let $\mathcal{Q}$ be a quantum permutation group on a set $X$, with universal action $u$. Then $u(f) = f \otimes 1$ is equivalent to $f$ being constant on the orbits of $\mathcal{Q}$.
\end{lemma}
\proof
Letting $f_i$ be the $i^\text{th}$ component of $f$, we have that
\begin{align*}
u(f) &= \sum_i f_i u(e_i) = \sum_{i,j} f_i e_j \otimes u_{ji} = \sum_{j} e_j \otimes \left(\sum_{i} f_i u_{ji}\right); \\
f \otimes 1 &= \sum_j e_j \otimes f_j 1.
\end{align*}
These two expressions are equal if and only if $\sum_i f_i u_{ji} = f_j 1$ for all $j \in X$. Multiplying both sides of this equation by $u_{jk}$ for some fixed $k \in X$, we see that a necessary condition for it holding is that $f_k u_{jk} = f_j u_{jk}$ for all $k \in X$. This implies that $f_k = f_j$ must hold for all $j,k$ such that $u_{jk} \ne 0$, i.e., that $f$ must be constant on the orbits of $\mathcal{Q}$.

On the other hand, if $f$ is constant on the orbits of $\mathcal{Q}$, then $f_i = f_j$ unless $u_{ji} = 0$. Therefore, $\sum_i f_i u_{ji} = \sum_i f_j u_{ji} = f_j \sum_i u_{ji} = f_j 1$ for all $j \in X$. Thus we have proven the lemma.\qeds

As a special case of the above, we see that the notion of transitivity of a quantum permutation group coincides with the more combinatorial definition of there being a single orbit.

\begin{corollary}
Let $\mathcal{Q}$ be a quantum permutation group acting on a finite set $X$. The following are equivalent:
\begin{enumerate}
\item $\mathcal{Q}$ acts transitively on $X$, i.e., the fixed point algebra $(\mathbb{C}^X)^{\mathcal{Q}}$ is trivial;
\item $\mathcal{Q}$ has only one orbit on $X$;
\item there exists $x \in X$ such that $u_{xy} \ne 0$ for all $y \in X$;
\item $u_{xy} \ne 0$ for all $x,y \in X$.
\end{enumerate}
\end{corollary}

We can also prove an analog of Lemma~\ref{lem:orbits2orbits} for orbitals as well. For this, we need to consider the tensor square of the representation $u$, which is the representation $u^{\otimes 2}$ of $\mathcal{Q}$ on $\mathbb{C}^X\otimes \mathbb{C}^X$ defined on the standard basis elements by:
\[u^{\otimes 2}(e_{i} \otimes e_{i'}) = \sum_{j, j'} (e_{j} \otimes e_{j'}) \otimes u_{ji}u_{j'i'}.\]
The space of fixed points of $u^{\otimes 2}$ is defined in the same manner as for $u$:
\[
\mathrm{Hom}(1,u^{\otimes 2}) = \{f \in \mathbb{C}^X \otimes \mathbb{C}^X: u^{\otimes 2}(f) = f \otimes 1\}.\]
Such a representation can be seen as the analog of the diagonal action, even though $u^{\otimes 2}$ is not in general an action. Nonetheless, we will show that $\mathrm{Hom}(1,u^{\otimes 2})$ is a selfadjoint subalgebra of $\mathbb{C}^X \otimes \mathbb{C}^X$, whose corresponding equivalence relation on $X\times X$ has the orbitals as classes.


\begin{lemma}\label{lem:orbitals2orbitals}
Let $\mathcal{Q}$ be a quantum permutation group on a set $X$, with universal action $u$. Then $u^{\otimes 2}(f) = f \otimes 1$ if and only if $f$ is constant on the orbitals of $\mathcal{Q}$.
\end{lemma}
\proof
We have that
\begin{align*}
u^{\otimes 2}(f) &= \sum_{x,x'} f_{xx'} u(e_{x} \otimes e_{x'}) = \sum_{x,x',y,y'} f_{xx'} (e_{y} \otimes e_{y'}) \otimes u_{y x}u_{y'x'} = \sum_{y, y'} (e_{y} \otimes e_{y'}) \otimes \left(\sum_{x, x'} f_{x x'} u_{yx}u_{y'x'}\right); \\
f \otimes 1 &= \sum_{y, y'} (e_{y} \otimes e_{y'}) \otimes f_{yy'}1.
\end{align*}
These two expressions are equal if and only if $\sum_{x, x'} f_{xx'} u_{yx}u_{y'x'} = f_{yy'} 1$ for all $y,y' \in X$. Multiplying both sides of this equation on the left by $u_{y z}$ and on the right by $u_{y'z'}$ for some $z, z' \in X$, we see that a necessary condition for it to hold is that $f_{zz'} u_{yz} u_{y'z'} = f_{yy'} u_{yz} u_{y'z'}$ for all $z, z' \in X$. This implies that $f_{zz'} = f_{yy'}$ whenever $u_{yz} u_{y'z'} \ne 0$, i.e., whenever $(y,y')$ and $(z, z')$ are in the same orbital of $\mathcal{Q}$. This proves that if $u^{\otimes 2}(f) = f \otimes 1$, then $f$ is constant on the orbitals of $\mathcal{Q}$.

Conversely, if $f$ is constant on the orbitals of $\mathcal{Q}$, then $f_{xx'} = f_{yy'}$ whenever $u_{yx}u_{y'x'} \ne 0$. Therefore,
\[\sum_{x, x'} f_{xx'} u_{yx}u_{y'x'} = \sum_{x, x'} f_{yy'} u_{yx}u_{y'x'} = f_{yy'} \sum_{x, x'} u_{yx}u_{y'x'} = f_{yy'} 1\]
as desired.\qeds

\subsection{A coherent configuration}\label{subsec:acc}

Given a group $\Gamma$ acting on a set $X$, it is well known that the orbitals of $\Gamma$ form a coherent configuration on $X$. Indeed, this was part of the original motivation for studying coherent configurations. Here we will show that the same holds for the orbitals of a quantum permutation group.

\begin{theorem}\label{thm:qcc}
Given a quantum permutation group $\mathcal{Q}$ acting on a set $X$, the orbitals of $\mathcal{Q}$ form a coherent configuration on $X$.
\end{theorem}
\proof
Let $\mathcal{R} = \{R_i : i \in \mathcal{I}\}$ for some index set $\mathcal{I}$ be the set of orbitals of $\mathcal{Q}$. It is clear from the definition that the orbitals partition the set $X \times X$. We must show that $\mathcal{R}$ satisfies Conditions 1--3 of the definition of a coherent configuration.

If $x \ne y$, then $u_{xx}u_{xy} = 0$ and thus $(x,x) \not\sim_2 (x,y)$ whenever $x \ne y$. Therefore every orbital of $\mathcal{Q}$ is either contained in or disjoint from the diagonal of $X \times X$. This implies that Condition 1 is satisfied.

Suppose that $(x,x') \sim_2 (y,y')$, i.e., that $u_{xy}u_{x'y'} \ne 0$. Then $0 \ne (u_{xy}u_{x'y'})^* = u_{x'y'}u_{xy}$ and thus $(x',x) \sim_2 (y',y)$. By symmetry we have that $(x,x') \sim_2 (y,y')$ if and only if $(x',x) \sim_2 (y',y)$, which implies that Condition 2 is met.

Let $i,j,k \in \mathcal{I}$ and suppose that $(x,z), (x',z') \in R_k$. Then $u_{xx'}u_{zz'} \ne 0$. Let $S = \{y : (x,y) \in R_i \ \& \ (y,z) \in R_j\}$ and $S' = \{y' : (x',y') \in R_i \ \& \ (y',z') \in R_j\}$. We aim to show that $|S| = |S'|$. We have that
\begin{align*}
|S| u_{xx'}u_{zz'} &= u_{xx'} \left(\sum_{y \in S} 1 \right) u_{zz'} = u_{xx'} \left(\sum_{y \in S} \sum_{y' \in X} u_{yy'} \right) u_{zz'} = \sum_{y \in S} \sum_{y' \in X} u_{xx'} u_{yy'} u_{zz'} \\
&= \sum_{y \in S} \sum_{y' \in S'} u_{xx'} u_{yy'} u_{zz'} \ \ \text{\Big(here we used $y' \not\in S'$ $\Rightarrow$ $u_{xx'}u_{yy'} = 0$ or $u_{yy'}u_{zz'} = 0$\Big)} \\
&= \sum_{y' \in S'} \sum_{y \in S} u_{xx'} u_{yy'} u_{zz'} = u_{xx'} \left(\sum_{y' \in S'} \sum_{y \in S} u_{yy'}\right) u_{zz'} = u_{xx'} \left(\sum_{y' \in S'} \sum_{y \in X} u_{yy'}\right) u_{zz'} \\
&= |S'| u_{xx'}u_{zz'}.
\end{align*}
This implies that the constants $p_{ij}^k$ exist and Condition 3 is met. Therefore, the orbitals $\mathcal{R}$ form a coherent configuration.\qeds

Due the to above theorem, we give the following definition:

\begin{definition}
Let $X$ be a graph with quantum automorphism group $\qut(X)$. We define the \emph{quantum orbital configuration/algebra} of $X$ as the coherent configuration/algebra produced by the orbitals of $\qut(X)$ acting on $V(X)$.
\end{definition}

We will similarly refer to the coherent configuration/algebra produced by the orbitals of $\aut(X)$ as the \emph{orbital configuration/algebra} of $X$. We note that the orbital configuration is a (possibly trivial) refinement of the quantum orbital configuration or, equivalently, the quantum orbital algebra is a subalgebra of the orbital algebra. This follows from the fact that $\mathrm{Aut}(X)$ is a quantum subgroup of $\mathrm{Qut}(X)$. More concretely, for any $\sigma \in \aut(X)$ the mapping $u_{x\sigma(x)} \mapsto 1$ and $u_{xy} \mapsto 0$ for all $y \ne \sigma(x)$ gives a representation of the quantum symmetry algebra of $X$. Thus if $\sigma(x) = y$ and $\sigma(x') = y'$, then $u_{xy}u_{x'y'} \ne 0$ since its image is nonzero in this representation. Therefore, if $(x,x')$ and $(y,y')$ are in the same (classical) orbital of $X$, they are in the same quantum orbital as well.



We remark here that the coherent configuration of a graph can ``detect" many properties of the graph. More precisely, the classes of the configuration can detect properties of the pairs of vertices they correspond to. For instance, if $X$ is a graph with adjacency matrix $A$, then for all $\ell \in \mathbb{N}$ the power $A^\ell$ of $A$ is in the coherent algebra of $X$, and the entries of $A^\ell$ record the number of walks of length $\ell$ between pairs of vertices. Thus if two pairs of vertices have a different number of walks of length $\ell$ between them, then they must be in different classes of the coherent configuration (since $A^\ell$ must lie in the span of the characteristic matrices of the classes). It is not too hard to see that this also means that pairs of vertices at different distances cannot be in the same class, and one can similarly find many other properties of pairs of vertices that are detected by the coherent configuration of the graph. For instance, this implies the adjacency matrix of the $d$-distance graph of $X$ is contained in its coherent algebra for all $d$. Since this holds for the coherent configuration of $X$, it also holds for the quantum orbital configuration of $X$. It is an interesting question what additional properties are detected by the latter configuration.

We cannot help but mention that Higman introduced coherent algebras in order to `do group theory without groups', and here we are constructing coherent algebras from quantum groups, which are another approach to studying group theory without groups.

\subsection{Commutation properties}\label{subsec:commprops}

In the classical case, the permutations in the automorphism group of a graph can be represented as permutation matrices. These permutation matrices are exactly those that commute with the adjacency matrix of the graph, but they will also commute with the characteristic matrices of the orbitals of the automorphism group. In fact, the orbital algebra of $X$ is exactly the commutant of $\aut(X)$ when considered as a subgroup of the permutation matrices. Here, we will prove the quantum analog of this fact.

\begin{theorem}\label{thm:qcommute}
Let $\mathcal{Q}$ be a quantum permutation group acting on a set $X$, let $U = (u_{xy})_{x,y \in X}$ be the magic unitary defining $\mathcal{Q}$, and let $M$ be a scalar matrix with index set $X$. Then $MU = UM$ if and only if $M$ is in the orbital algebra of $\mathcal{Q}$, i.e., $M$ is constant on the orbitals of $\mathcal{Q}$.
\end{theorem}
\proof
Let $M = (m_{xy})_{x,y \in X}$. Also, let $R_i$ for $i \in \mathcal{I}$ be the orbitals of $\mathcal{Q}$. We can write the entries of $MU$ as follows:
\begin{align}
(MU)_{xz} &= \sum_{y \in X} m_{xy}u_{yz} = \left(\sum_{y' \in X} u_{xy'}\right)\left(\sum_{y \in X} m_{xy}u_{yz}\right) \\
&= \sum_{y,y' \in X} m_{xy} u_{xy'}u_{yz} = \sum_{y,y' : (x,y) \sim_2 (y',z)} m_{xy} u_{xy'}u_{yz} \\
&= \sum_{i \in \mathcal{I}}\sum_{y,y': (x,y),(y',z) \in R_i} m_{xy} u_{xy'}u_{yz}\label{eq:MU}
\end{align}
Similarly, we have
\begin{align}
(UM)_{xz} &= \sum_{y' \in X} m_{y'z}u_{xy'} = \left(\sum_{y' \in X} m_{y'z}u_{xy'}\right)\left(\sum_{y \in X} u_{yz}\right) \\
&= \sum_{y,y' \in X} m_{y'z} u_{xy'}u_{yz} = \sum_{y,y' : (x,y) \sim_2 (y',z)} m_{y'z} u_{xy'}u_{yz} \\
&= \sum_{i \in \mathcal{I}}\sum_{y,y': (x,y),(y',z) \in R_i} m_{y'z} u_{xy'}u_{yz}\label{eq:UM}
\end{align}
Suppose that $M$ is constant on the orbitals of $\mathcal{Q}$ and that $m_{vw} =: m_i$ for all $(v,w) \in R_i$. Then the expressions in~(\ref{eq:MU}) and~(\ref{eq:UM}) both become
\[\sum_{i \in \mathcal{I}} m_i \left(\sum_{y,y': (x,y),(y',z) \in R_i} u_{xy'}u_{yz}\right),\]
and thus $MU = UM$.

Conversely, if $MU = UM$, then the expressions in~(\ref{eq:MU}) and~(\ref{eq:UM}) must be equal for all $x,z \in X$. Suppose that $m_{xv} \ne m_{wz}$ for some $(x,v), (w,z) \in R_j$ for some $j \in \mathcal{I}$. Then multiplying the expression in~(\ref{eq:MU}) on the left by $u_{xw}$ and on the right by $u_{vz}$ we obtain
\[m_{xv}u_{xw}u_{vz}.\]
Doing the same to the expression in~(\ref{eq:UM}), we obtain
\[m_{wz}u_{xw}u_{vz}.\]
Since $MU = UM$ by assumption, we have that these two expressions must be equal. Furthermore, $u_{xw}u_{vz} \ne 0$ since $(x,v)$ and $(w,z)$ are in the same orbital. Thus we have that $m_{xv} = m_{wz}$, a contradiction to our assumption. Therefore, if $MU = UM$, then $M$ must be constant on the orbitals of $\mathcal{Q}$, i.e., $M$ must be contained in the orbital algebra of $\mathcal{Q}$.\qeds

%

The purpose of the above theorem is to provide more information about the quantum automorphism group of a graph. However, one might argue that it could be difficult or impossible to determine the quantum orbitals of a given graph, and so the above may not be of much use. But, as we mentioned previously, one can efficiently compute the coherent configuration of a graph. In the next section we will see that this is a useful tool for studying the quantum automorphism group of a graph.

\subsection{Using the coherent algebra of a graph}\label{subsec:useca}

Recall from the end of Section~\ref{subsec:acc}, we noted that the quantum orbital algebra of a graph $X$ is a subalgebra of the orbital algebra of $X$, i.e., the orbital configuration of $X$ is a refinement of the quantum orbital configuration of $X$. On the other hand, the quantum orbital algebra of a graph must contain its adjacency matrix, since $(x,x') \sim_2 (y,y')$ (i.e.~$u_{xy}u_{x'y'} \ne 0$) is not possible if one of the pairs form an edge and the other does not. Therefore, the quantum orbital algebra of a graph $X$ must contain the coherent algebra of $X$, and thus the classes of the coherent configuration of $X$ must be unions of the orbitals of $\qut(X)$. From this we obtain the following two corollaries:


\begin{corollary}\label{cor:cacommute}
Let $X$ be a graph with adjacency matrix $A$. Then a magic unitary $U$ that commutes with $A$ also commutes with every element of the coherent algebra of $X$.
\end{corollary}


\begin{corollary}\label{cor:extrarels}
Let $X$ be a graph with adjacency matrix $A$. If $U$ is a magic unitary that commutes with $A$, then $u_{xy}u_{x'y'} = 0$ unless $(x,x')$ and $(y,y')$ are in the same class of the coherent configuration of $X$.
\end{corollary}

Using the second corollary above, we can show that almost all graphs have trivial quantum automorphism group.

\begin{theorem}\label{thm:trivial}
Let $X$ be a random graph on $n$ vertices. The probability that $X$ has nontrivial quantum automorphism group goes to zero as $n$ goes to infinity.
\end{theorem}
\proof
Let $U$ be the magic unitary defining $\qut(X)$. It is known (it follows from Theorem 4.1 of~\cite{babaikucera} for instance) that almost all graphs have coherent algebras equal to the full matrix algebra, i.e., their coherent configuration is the partition of $V(X) \times V(X)$ into singletons. This implies that the orbitals of $\qut(X)$ are singletons. Since the orbits are just the orbitals contained in the diagonal (where when we identify the vertex set of $X$ with the diagonal of $X \times X$), we have that the orbits of $\qut(X)$ are singletons as well. Therefore, $u_{xy} = 0$ unless $x = y$, and this implies $u_{xx} = 1$ for all $x \in V(X)$ by the definition of magic unitary. Therefore the magic unitary $U$ defining $\qut(X)$ is the identity matrix for almost all graphs $X$, so thus they have trivial quantum automorphism group.\qeds

Though the above result is not surprising (it is well known that it holds for classical automorphisms), it is not immediately obvious how to prove this in the quantum case. Thus the above theorem exhibits the usefulness of coherent configurations/algebras in the study of quantum permutation groups. In Section~\ref{sec:discuss}, we will discuss a possible application of Theorem~\ref{thm:qcommute} for showing that certain graphs have no quantum symmetry.

\section{Quantum Isomorphism}\label{sec:qiso}
 
In Section~\ref{subsec:isogame}, we introduced the notions of quantum tensor isomorphism and quantum commuting isomorphism. Here we will define a notion of quantum isomorphism that fits naturally with the already existing notion of quantum automorphism.

\begin{definition}
We say that graphs $X$ and $Y$ with adjacency matrices $A$ and $B$ are \emph{quantum isomorphic}, and write $X \cong_q Y$, if there exists a Hilbert space $\mathcal{H}$, a magic unitary $U = (u_{xy})$ with $u_{xy}\in \mathcal{B}(\mathcal{H})$ for all $x \in V(X), y \in V(Y)$ such that $AU = UB$ or, equivalently, $u_{xy}u_{x'y'} = 0$ if $\rel(x,x') \ne \rel(y,y')$.
\end{definition}

Because of the correspondence between $C^*$-algebras and algebras of bounded linear operators on a Hilbert space, the above is easily seen to be equivalent to the definition of quantum commuting isomorphism given by Theorem~\ref{thm:qciso}, except without the requirement of a tracial state. Therefore, $X \cong_{qc} Y$ implies $X \cong_q Y$ for any graphs $X$ and $Y$. We have already mentioned that $X \cong_{qt} Y \Rightarrow X \cong_{qc} Y$, and so we have the following chain of implications:
\[X \cong Y \ \Rightarrow \ X \cong_{qt} Y \ \Rightarrow \ X \cong_{qc} Y \ \Rightarrow \ X \cong_{q} Y.\]
The fact that the first two implications cannot be reversed was proven in~\cite{qiso1}. In Section~\ref{subsec:qcequalsq}, we will show that the last implication can be reversed, thus quantum isomorphism and quantum commuting isomorphism are the same relation.

It also follows from the above that the non-isomorphic but quantum tensor isomorphic graphs constructed in~\cite{qiso1} provide us with examples of non-isomorphic but quantum isomorphic graphs. We do not contribute anything more on this front, but we will see later that we can use these examples to construct graphs which are quantum vertex transitive but not vertex transitive, answering a question from the quantum permutation group literature. First, we will show that if two graphs are quantum isomorphic, then there is an isomorphism of their quantum orbital algebras that maps the adjacency matrix of one to that of the other. This result is inspired by results in an upcoming work~\cite{qiso2} of Varvitsiotis and the second and third authors, in which they show that a certain semidefinite relaxation of quantum tensor/commuting isomorphism is equivalent to the existence of an isomorphism of the coherent algebras of two graphs which maps the adjacency matrix of one to the other.

%

\subsection{The Haar state}\label{subsec:haar}

In this section, we will compute the value of the Haar state of a quantum permutation group on products of pairs of generators. We will see that the value depends only on the sizes of the orbitals of the quantum permutation groups.

\begin{theorem}\label{thm:haar}
Let $\mathcal{Q}$ be a quantum permutation group on a set $X$. Let $U = (u_{xy})$ be the magic unitary defining $C(\Q)$ and let $h$ denote the Haar state on $C(\mathcal{Q})$. If $R_1, \ldots, R_m \subseteq X \times X$ are the orbitals of $\Q$ on $X$, then
\[
h(u_{xy}u_{x'y'}) = 
\begin{cases}
1/|R_i| &\text{if } (x,x'),(y,y') \in R_i; \\
0 &\text{if } (x,x') \not\sim_2 (y,y').
\end{cases}
\]
\end{theorem}
\proof
Note that the $(x,x') \not\sim_2 (y,y')$ case is trivial since in this case $u_{xy}u_{x'y'} = 0$ by definition.

So suppose that $(x,x'),(y,y'),(z,z') \in R_i$. We will show that $h(u_{xy}u_{x'y'}) = h(u_{xz}u_{x'z'})$, i.e., changing the second indices to another pair of vertices in the same orbital does not change the value of $h$. Recall from Section~\ref{subsec:qperms} that $(h \otimes \text{id})\Delta = (\text{id} \otimes h)\Delta = h(\cdot)1$. Therefore, we have that
\begin{align*}
h(u_{xz}u_{x'z'})1 &= (h \otimes \text{id})\Delta(u_{xz}u_{x'z'}) = (h \otimes \text{id})\Delta(u_{xz})\Delta(u_{x'z'}) \\
&= (h \otimes \text{id})\left(\left(\sum_{w \in X} u_{xw} \otimes u_{wz}\right)\left(\sum_{w' \in X} u_{x'w'} \otimes u_{w'z'}\right)\right) \\
&= (h \otimes \text{id})\left(\sum_{w,w' \in X} u_{xw}u_{x'w'} \otimes u_{wz}u_{w'z'}\right) \\
&= (h \otimes \text{id})\left(\sum_{(w,w') \in R_i} u_{xw}u_{x'w'} \otimes u_{wz}u_{w'z'}\right) \\
&= \sum_{(w,w') \in R_i} h(u_{xw}u_{x'w'}) u_{wz}u_{w'z'}.
\end{align*}
Multiplying this equation on the left by $u_{yz}$ and on the right by $u_{y'z'}$, we obtain
\[h(u_{xz}u_{x'z'}) u_{yz}u_{y'z'} = h(u_{xy}u_{x'y'}) u_{yz}u_{y'z'}.\]
Since $(y,y'),(z,z') \in R_i$, we have that $u_{yz}u_{y'z'} \ne 0$, and therefore $h(u_{xz}u_{x'z'}) = h(u_{xy}u_{x'y'})$. This shows that for $(x,x') \in R_i$, we have that $h(u_{xw}u_{x'w'})$ is constant for all $(w,w') \in R_i$ (and is zero otherwise). 

Now, again fixing $(x,x') (y,y') \in R_i$, we have that
\[1 = h(1) = h\left(\sum_{w,w' \in X} u_{xw}u_{x'w'}\right) = \sum_{w,w' \in X} h(u_{xw}u_{x'w'}) = \sum_{(w,w') \in R_i} h(u_{xw}u_{x'w'}) = |R_i| h(u_{xy}u_{x'y'}).\]
Thus $h(u_{xy}u_{x'y'}) = 1/|R_i|$, and $(x,x'),(y,y')$ were arbitrary elements of $R_i$ so we are done.\qeds

As a consequence of the above and Remark~\ref{rem:orbitalsagree}, we have the following:
\begin{corollary}\label{cor:haar}
Let $\mathcal{Q}$ be a quantum permutation group on a set $X$. Let $U = (u_{xy})$ be the magic unitary defining $C(\Q)$ and let $h$ denote the Haar state on $C(\mathcal{Q})$. If $O_1, \ldots, O_m \subseteq X$ are the orbits of $\Q$ on $X$, then
\[
h(u_{xy}) = 
\begin{cases}
1/|O_i| &\text{if } x, y \in O_i; \\
0 &\text{if } x \not\sim_1 y.
\end{cases}
\]
\end{corollary}


\subsection{Quantum isomorphism and quantum commuting isomorphism are the same}\label{subsec:qcequalsq}

In the beginning of Section~\ref{sec:qiso}, we noted that $X \cong_{qc} Y$ implies that $X \cong_q Y$ for any graphs $X$ and $Y$. Here we will show that this implication can be reversed, thus proving that these two relations are equivalent.

\begin{theorem}
Let $X$ and $Y$ be graphs. Then $X \cong_q Y$ if and only if $X \cong_{qc} Y$.
\end{theorem}
\proof
We only need to show that if $X \cong_q Y$, then $X \cong_{qc} Y$. So suppose that the former holds. By taking complements if necessary, we may assume that $X$ is connected. It follows from Corollary~\ref{cor:qcciso} that $Y$ is also connected. Let $Z = X \cup Y$ be the disjoint union of $X$ and $Y$. Also, let $U = (u_{wz})$ for $w,z \in V(Z)$ be the magic unitary defining $C(\qut(Z))$. We will show that $u_{xy}u_{x'x''} = 0$ for all $y \in V(Y)$ and $x,x',x'' \in V(X)$.

Since $X$ is connected, the distance in $Z$ between $x,x' \in V(X)$ is finite. On the other hand, there is no path between $y$ and $x''$ in $Z$. This implies that $(x,x')$ and $(y,x'')$ cannot be in the same class of the coherent configuration of $Z$ (see the discussion at the end of Section~\ref{subsec:acc}). Since the classes of the quantum orbital configuration are refinements of the classes of the coherent configuration, we see that $(x,x')$ and $(y,x'')$ cannot be in the same orbital of $\qut(Z)$. Therefore, $u_{xy}u_{x'x''} = 0$ as desired.

Now, for each $x \in V(X)$, define the projection $p_x = \sum_{y \in V(Y)} u_{xy}$. We will show that $p_x = p_{x'}$ for all $x,x' \in V(X)$. From the above, we have that
\[p_x(1 - p_{x'}) = p_x \left(\sum_{x'' \in V(X)} u_{x'x''}\right) = 0.\]
So $p_x = p_xp_{x'}$, and similarly we can show that $p_{x'} = p_xp_{x'}$. Thus $p_x = p_{x'}$ as desired. Similarly, we can define $p_y = \sum_{x \in V(X)} u_{xy}$, and these will all be equal. Next we want to show that $p_x = p_y$. We have that
\[|V(X)|p_x = \sum_{x \in V(X)} p_x = \sum_{x \in V(X), y \in V(Y)} u_{xy} = \sum_{y \in V(Y)} p_y = |V(Y)|p_y.\]
Since $X$ and $Y$ are quantum isomorphic, we have that $|V(X)| = |V(Y)|$ and so $p_x = p_y$. Since all of these projections are equal, we will refer to them simply as $p$ from now on.

Next, we want to show that $p \ne 0$. Since $X \cong_q Y$, we have that there are operators $v_{xy}$ for $x \in V(X)$, $y \in V(Y)$ on some Hilbert space such that the matrix $V = (v_{xy})$ is a magic unitary and $v_{xy}v_{x'y'} = 0$ whenever $\rel(x,x') \ne \rel(y,y')$. Then it is easy to see that mapping $u_{xy}$ to $v_{xy}$ for $x \in V(X)$, $y \in V(Y)$ and mapping all other $u_{wz}$ to zero is a representation of $C(\qut(X))$. Moreover, in this representation, the element $p$ is mapped to $\sum_{y \in V(Y)} v_{xy} = 1$. Since $p$ is mapped to a nonzero element in this representation, we have that $p \ne 0$ in $C(\qut(Z))$. Note that this implies that for any $x \in V(X)$, there exists some $y \in V(Y)$ such that $x$ and $y$ are in the same orbit of $\qut(Z)$.

Now let $\mathcal{A}$ be the $C^*$-subalgebra of $C(\qut(Z))$ generated by the elements $u_{xy}$ for $x \in V(X)$, $y \in V(Y)$. Note that $p$ is the identity in $\mathcal{A}$, since
\[u_{xy}p = u_{xy} \sum_{y' \in V(Y)} u_{xy'} = u_{xy}^2 = u_{xy},\]
and similarly $pu_{xy} = u_{xy}$. This means that the matrix $\hat{U} = (u_{xy})$ for $x \in V(X)$, $y \in V(Y)$ is a magic unitary over $\mathcal{A}$. Moreover, we have that $u_{xy}u_{x'y'} = 0$ if $\rel(x,x') \ne \rel(y,y')$, since this was true in $C(\qut(Z))$. Thus, $\hat{U}$ is a magic unitary such that $A\hat{U} = \hat{U}B$ where $A$ and $B$ are the adjacency matrices of $X$ and $Y$ respectively. This only shows that $X \cong_q Y$, which we had already assumed, and so we still need to construct a tracial state on $\mathcal{A}$.

Let $h$ be the Haar state on $C(\qut(Z))$, and recall that this is tracial, i.e., $h(ab) = h(ba)$ for all $a,b \in C(\qut(Z))$. It is easy to see that the restriction of $h$ to $\mathcal{A}$ satisfies all of the conditions of being a tracial state on $\mathcal{A}$, except that it may not evaluate to one on the identity $p$. Recall from above that for any $x \in V(X)$ there is some $y \in V(Y)$ such that $x$ and $y$ are in the same orbit of $\qut(Z)$. Thus, by definition of $p$ and Corollary~\ref{cor:haar} we have that $h(p) > 0$, and so we can define $\hat{h}(\cdot) = h(\cdot)/h(p)$. This is our tracial state on $\mathcal{A}$, and so by Theorem~\ref{thm:qciso} and Remark~\ref{rem:faithful} we are done.\qeds

Note that in the above, the only place where we used the fact that $X \cong_{qc} Y$ was to assume without loss of generality that $X$ and $Y$ are both connected and to show that $p \ne 0$. The latter happens if and only if there exists $x \in V(X)$ and $y \in V(Y)$ such that $u_{xy} \ne 0$. Therefore, we immediately have the following:

\begin{theorem}\label{thm:iso2orb}
Let $X$ and $Y$ be connected graphs. Then $X \cong_{qc} Y$ if and only if there exists $x \in V(X)$ and $y \in V(Y)$ that are in the same orbit of $\qut(X \cup Y)$.
\end{theorem}

Since any graph is either connected or has a connected complement, and connected graphs can only be quantum isomorphic to connected graphs, the above reduces the question of quantum isomorphism to the determination of the orbits of a quantum permutation group, which is exactly analogous to the classical case.

\subsection{Isomorphisms of coherent algebras}\label{subsec:isoca}

Given coherent algebras $\mathcal{C}$ and $\mathcal{C}'$, an isomorphism $\Phi: \mathcal{C} \to \mathcal{C}'$ is a linear bijection which, for any $M,N \in \mathcal{C}$, satisfies
\begin{itemize}
\item $\Phi(MN) = \Phi(M)\Phi(N)$;
\item $\Phi(M^*) = \Phi(M)^*$;
\item $\Phi(M \circ N) = \Phi(M) \circ \Phi(N)$, where $\circ$ denotes Schur product;
\item $\Phi(I) = I$ and $\Phi(J) = J$.
\end{itemize}

Consider the coherent configurations $\mathcal{R} = \{R_i : i \in \mathcal{I}\}$ and $\mathcal{R}' = \{R'_i : i \in \mathcal{I}'\}$ corresponding to $\mathcal{C}$ and $\mathcal{C}'$. Suppose that $\mathcal{R}$ and $\mathcal{R}'$ have intersection numbers $p_{ij}^k$ for $i,j,k \in \mathcal{I}$ and $q_{ij}^k$ for $i,j,k \in \mathcal{I}'$ respectively. It is well known and not too difficult to see that an isomorphism from $\mathcal{C}$ to $\mathcal{C}'$ is equivalent to a bijection $f: \mathcal{I} \to \mathcal{I}'$ such that $p_{ij}^k = q_{f(i)f(j)}^{f(k)}$ for all $i,j,k \in \mathcal{I}$. The corresponding isomorphism $\Phi$ will then map the characteristic matrix of $R_i$ to that of $R'_{f(i)}$ for all $i \in \mathcal{I}$.

It is obvious that isomorphic graphs have isomorphic orbital algebras. The isomorphism is given by conjugation with a permutation matrix encoding the isomorphism of the two graphs. However, the converse does not hold: there are graphs with isomorphic orbital algebras which are not isomorphic themselves. An example would be the Paley and Peisert graphs of the same order.

Here we show that quantum isomorphic graphs have isomorphic quantum orbital algebras, in analogy with the fact for isomorphic graphs.

\begin{theorem}\label{thm:qcciso}
Suppose that $X$ and $Y$ are quantum isomorphic graphs with adjacency matrices $A$ and $B$. Then there exists an isomorphism $\Phi$ of their quantum orbital algebras such that $\Phi(A) = B$.
\end{theorem}
\proof
Let $\mathcal{R} = \{R_i : i \in \mathcal{I}\}$ and $\mathcal{R}' = \{R'_i : i \in \mathcal{I}'\}$ be the quantum orbital configurations of $X$ and $Y$ respectively, with intersection numbers $p_{ij}^k$ and $q_{ij}^k$. Let $V = (v_{xy})$ be the magic unitary witnessing the quantum isomorphism of $X$ and $Y$. Assume that $V(X)$ and $V(Y)$ are disjoint and define a relation $\sim_v$ on $(V(X) \times V(X)) \cup (V(Y) \times V(Y))$ by $(x,x') \sim_v (y,y')$ if $v_{xy}v_{x'y'} \ne 0$. We will think of this relation as defining a bipartite graph $W$ with parts $V(X) \times V(X)$ and $V(Y) \times V(Y)$. We will show that this relation gives rise to a bijection $f: \mathcal{I} \to \mathcal{I}'$ that provides us with the required isomorphism.

First, we will use our quantum isomorphism of $X$ and $Y$ to construct a magic unitary that commutes with $A$. Let $U = (u_{yy'})$ for $y,y' \in V(Y)$ be the magic unitary defining the quantum symmetry algebra of $Y$. Consider the matrix $\hat{V} = (\hat{v}_{xx'})$ for $x,x' \in V(X)$ defined as
\[\hat{v}_{xx'} = \sum_{y,y' \in V(Y)} v_{xy} \otimes u_{yy'} \otimes v_{x'y'}.\]
Intuitively, this is capturing the idea of composing an isomorphism from $X$ to $Y$ with an automorphism of $Y$ and then followed by an isomorphism from $Y$ back to $X$. It is easy to see that $\hat{v}_{xx'} = \hat{v}^*_{xx'}$ for all $x \in x' \in V(X)$. Moreover,
\[\hat{v}_{xx'}^2 = \sum_{y,y',w,w' \in V(Y)} v_{xy}v_{xw} \otimes u_{yy'}u_{ww'} \otimes v_{x'y'}v_{x'w'} = \sum_{y,y'} v_{xy} \otimes u_{yy'} \otimes v_{x'y'} = \hat{v}_{xx'},\]
since $v_{xy}v_{xw} = 0$ unless $w = y$ and $v_{x'y'}v_{x'w'} = 0$ unless $w' = y'$. We also have that
\begin{align*}
\sum_{x' \in V(X)} \hat{v}_{xx'} &= \sum_{x' \in V(X), y,y' \in V(Y)} v_{xy} \otimes u_{yy'} \otimes v_{x'y'} \\
&= \sum_{y,y' \in V(Y)} v_{xw} \otimes u_{yy'}\otimes \left(\sum_{x'} v_{x'w}\right) \\
&= \sum_{y \in V(Y)} v_{xy} \otimes \left(\sum_{y'} u_{yy'}\right) \otimes 1 \\
&= \sum_{y \in V(Y)} v_{xy} \otimes 1 = 1,
\end{align*}
and similarly $\sum_x \hat{v}_{xx'} = 1$. Thus $\hat{V}$ is a magic unitary.

Now suppose that $\rel(x',x'') \ne \rel(x^*,x^{**})$. Then
\[\hat{v}_{x'x^*}\hat{v}_{x''x^{**}} = \sum_{y',y^*,y'',y^{**} \in V(Y)} v_{x'y'}v_{x''y''} \otimes u_{y'y^*}u_{y''y^{**}} \otimes v_{x^*y^*}v_{x^{**}y^{**}} = 0,\]
since any nonzero term in the sum would require $\rel(x',x'') = \rel(y',y'') = \rel(y^*,y^{**}) = \rel(x^*,x^{**})$, a contradiction. Therefore, $A \hat{V} = \hat{V}A$ as desired.

Now suppose that $(x_1,x'_1), (x_2,x'_2) \in V(X) \times V(X)$ and that there exist $(y_1,y_1'), (y_2,y'_2)$ in the \emph{same} quantum orbital of $Y$ such that both $(x_1,x'_1) \sim_v (y_1,y_1')$ and $(x_2,x'_2) \sim_v (y_2,y_2')$. In other words, we have that $v_{x_1y_1}v_{x'_1y'_1} \ne 0$, $v_{x_2y_2}v_{x'_2y'_2} \ne 0$, and $u_{y_1y_2}u_{y'_1y'_2} \ne 0$. Then
\begin{align*}
\hat{v}_{x_1x_2}\hat{v}_{x'_1x'_2} = \sum_{y,y',w,w' \in V(Y)} v_{x_1y}v_{x'_1w} \otimes u_{yy'}u_{ww'} \otimes v_{x_2y'}v_{x'_2w'},
\end{align*}
and thus
\[\left(v_{x_1y_1}\otimes 1 \otimes v_{x_2y_2}\right)\left(\hat{v}_{x_1x_2}\hat{v}_{x'_1x'_2}\right)\left(v_{x'_1y'_1} \otimes 1 \otimes v_{x'_2y'_2}\right) = v_{x_1y_1}v_{x'_1y'_1} \otimes u_{y_1y_2}u_{y'_1y'_2} \otimes v_{x_2y_2}v_{x'_2y'_2} \ne 0.\]
Therefore we have that $\hat{v}_{x_1x_2}\hat{v}_{x'_1x'_2} \ne 0$ and so $(x_1,x'_1)$ and $(x_2,x'_2)$ must be in the same quantum orbital of $X$.

So we have shown that if we fix a quantum orbital, say $R'_i$, of $Y$, then the elements of $R'_i$ can be adjacent in $W$ only to elements of $R_j$ for some fixed $j$. It can similarly be shown that the same holds with $X$ and $Y$ swapped. Moreover, every vertex of $B$ has at least one neighbor since $\sum_{y_1,y_2} v_{xy_1}v_{x'y_2} = \sum_{x_1,x_2} v_{x_1y}v_{x_2y'} = 1$ for all $x,x' \in V(X)$ and $y,y' \in V(Y)$. Thus there is a bijection $f : \mathcal{I} \to \mathcal{I}'$ such that the elements of $R_i$ are adjacent in $W$ only to elements of $R'_{f(i)}$, i.e., $v_{xy}v_{x'y'} = 0$ unless $(x,x') \in R_i$ and $(y,y') \in R'_{f(i)}$ for some $i \in \mathcal{I}$.

Now we must show that the function $f$ has the property that $p_{ij}^k = q_{f(i)f(j)}^{f(k)}$ for all $i,j,k \in \mathcal{I}$. Suppose that $(x,x') \in R_k$ for some $k \in \mathcal{I}$, and let $i,j \in \mathcal{I}$. Define $S = \{w \in V(X): (x,w) \in R_i, (w,x') \in R_j\}$, and note that $|S| = p_{ij}^k$. There must exist $y,y' \in V(Y)$ such that $v_{xy}v_{x'y'} \ne 0$ and furthermore we must have that $(y,y') \in R'_{f(k)}$. Define $S' = \{z \in V(Y): (y,z) \in R'_{f(i)}, (z,y') \in R'_{f(j)}\}$ and note that $|S'| = q_{f(i)f(j)}^{f(k)}$. Following the proof of Theorem~\ref{thm:qcc}, we see that
\begin{align*}
p_{ij}^kv_{xy}v_{x'y'} &= v_{xy}\left(\sum_{w \in S} 1\right)v_{x'y'} \\
&= v_{xy}\left(\sum_{w \in S, z \in V(Y)} v_{wz}\right)v_{x'y'} \\
&= v_{xy}\left(\sum_{w \in S, z \in S'} v_{wz}\right)v_{x'y'} \\
&= v_{xy}\left(\sum_{w \in V(X), z \in S'} v_{wz}\right)v_{x'y'} \\
&= q_{f(i)f(j)}^{f(k)}v_{xy}v_{x'y'}.
\end{align*}
Since $v_{xy}v_{x'y'} \ne 0$ by assumption, we have that $p_{ij}^k = q_{f(i)f(j)}^{f(k)}$ as desired.

Let $A^i$ and $B^i$ be the characteristic matrices of the relations $R_i$ and $R'_i$ respectively. Letting $\Phi$ be linear extension to $\text{span}\{A^i : i \in \mathcal{I}\}$ of the map taking $A^i$ to $B^{f(i)}$ gives the isomorphism of the coherent algebras corresponding to $\mathcal{R}$ and $\mathcal{R}'$. The fact that $\Phi(A_X) = A_Y$ is immediate from the fact that $v_{xy}v_{x'y'} = 0$ when $\rel(x,x') \ne \rel(y,y')$ and thus $f$ maps indices corresponding to edges of $X$ to indices corresponding to edges of $Y$ and vice versa.\qeds

As with the classical case, we do not expect that the converse of the above theorem holds, though we do not currently know a counterexample. Note that the above theorem implies that quantum isomorphic graphs have the same number of quantum orbits. This follows from the fact that any isomorphism $\Phi$ of coherent algebras satisfies $\Phi(I) = I$ by definition, and this corresponds to mapping the diagonal relations of one coherent configuration to the diagonal relations of the other (consider the characteristic matrices of the diagonal relations).

The above theorem gives us a necessary condition for two graphs to be quantum isomorphic. However, as with Theorem~\ref{thm:qcommute}, it may be difficult or impossible to make use of this since we may not be able to compute the quantum orbitals of a given graph. But as we did before, we can fall back on the coherent algebras of the graphs to find a more useable necessary condition.

\begin{corollary}\label{cor:qcciso}
Suppose that $X$ and $Y$ are quantum isomorphic graphs with adjacency matrices $A$ and $B$. Then there exists an isomorphism $\Phi$ of their coherent algebras such that $\Phi(A) = B$.
\end{corollary}
\proof
Let $\Phi$ be the isomorphism of the quantum coherent algebras of $X$ and $Y$ satisfying $\Phi(A) = B$ guaranteed by Theorem~\ref{thm:qcciso}. It is easy to see that the restriction $\hat{\Phi}$ of $\Phi$ to the coherent algebra of $X$ has as its image a coherent algebra. Moreover, since $\Phi(A) = B$, we have that this image must contain the coherent algebra of $Y$. If the image of $\hat{\Phi}$ is strictly larger than the coherent algebra of $Y$, then the image of the restriction of $\Phi^{-1}$ to the coherent algebra of $Y$ would necessarily be strictly smaller than the coherent algebra of $X$. However, this is a contradiction since this image must contain the coherent algebra of $X$ by the same reasoning as with the image of $\hat{\Phi}$. Therefore, $\hat{\Phi}$ is an isomorphism of the coherent algebras of $X$ and $Y$ such that $\hat{\Phi}(A) = B$.\qeds

As mentioned in the introduction, the existence of an isomorphism $\Phi$ of the coherent algebras of $X$ and $Y$ which maps $A$ to $B$ is a previously studied relation known as graph \emph{equivalence}~\cite{symmpowers}, and the map $\Phi$ is said to be an equivalence of $X$ and $Y$. It is also known that two graphs are equivalent if and only if they cannot be distinguished by the (2-dimensional) Weisfeiler-Leman algorithm. Since this algorithm runs in polynomial time, this provides us with an efficiently testable necessary condition for quantum isomorphism.

Graphs $X$ and $Y$ being equivalent implies further necessary conditions. For instance, it is known that such graphs must be cospectral, not just with respect to their adjacency matrices, but with respect to several other matrices often associated to graphs such as the Laplacian, signless Laplacian, normalized Laplacian, etc. In fact, if $\Phi$ is an equivalence of $X$ and $Y$, then for any Hermitian matrix $M$ in the coherent algebra of $X$, the matrix $\Phi(M)$ will be cospectral to $M$. Equivalent graphs must also have the same diameter and radius, and share many other structural properties~\cite{furer2017}. Also, an equivalence of graphs $X$ and $Y$ maps the adjacency matrix of the $d$-distance graph of $X$ to that of $Y$, and this further implies that these distance graphs are equivalent. Similarly, quantum isomorphic graphs have quantum isomorphic $d$-distance graphs. This last fact follows from similar techniques as those presented here, but we do not give a full proof.

Recall that the relation of graph equivalence has been shown to be equivalent \emph{doubly nonnegative isomorphism}, a semidefinite relaxation of quantum tensor/commuting isomorphism defined in~\cite{qiso2}. The above corollary shows that this is also a relaxation of quantum isomorphism. They also show in~\cite{qiso2} that doubly nonnegative isomorphism is not equivalent to quantum tensor/commuting isomorphism. We remark here that the same proof shows that it is not equivalent to quantum isomorphism as defined here.

We note here that the notion of quantum isomorphism can be used to find graphs that are quantum vertex transitive but not vertex transitive. First, one should observe that there exist graphs $X,Y$, one of which is vertex transitive, that are quantum isomorphic but not isomorphic. This will be proved later in this section. Granted this, one can take their disjoint union, which will be quantum vertex transitive but not vertex transitive.


\subsection{Linear Binary Constraint Systems}\label{subsec:LBCS}

A linear binary constraint system (LBCS) is simply a system of linear equations (constraints) over $\mathbb{F}_2$. More formally, an LBCS $\mathcal{F}$ consists of a family of binary variables $x_1, \ldots, x_n$ and constraints $C_1, \ldots, C_m$, where each $C_\ell$ takes the form $\sum_{x_i \in S_\ell} x_i = b_\ell$ for some $S_\ell \subseteq \{x_1, \ldots, x_n\}$ and $b_\ell \in \mathbb{F}_2$. An LBCS is said to be satisfiable if there is an assignment of values from $\mathbb{F}_2$ to the variable $x_i$ such that every constraint $C_\ell$ is simultaneously satisfied.

\begin{align}\label{eqn:BCS}
x_1 + x_2 + x_3 &= 0 \quad &x_1 + x_4 + x_7 = 0 \nonumber\\
x_4 + x_5 + x_6 &= 0 \quad &x_2 + x_5 + x_8 = 0 \\
x_7 + x_8 + x_9 &= 0 \quad &x_3 + x_6 + x_9 = 1\nonumber
\end{align}
where we stress again that the equations are over $\mathbb{F}_2$. It is not too difficult to see that the LBCS above is not satisfiable. This is because every variable appears in exactly two constraints and thus adding up all of them we obtain $0 = 1$.

There is an important alternative formulation of LBCS's that we will briefly mention. One can replace each constraint $\sum_{x_i \in S_\ell} x_i = b_\ell$ with $\prod_{x_i \in S_\ell} x_i = (-1)^{b_\ell}$, and allow the variables to take values in $\{-1,1\}$ to obtain an equivalent system of equations which we will call the multiplicative form of the LBCS. This form is used in the work of Cleve and Mittal~\cite{clevemittal}, where they construct a nonlocal game based on LBCS's (actually they do it even for nonlinear binary constraint systems). In the game, Alice is asked a constraint and Bob a variable in her constraint. In order to win, Alice must respond with an assignment of values to all of the variables in her constraint such that it is satisfied, and Bob must assign his variable a value so that it agrees with Alice. Cleve and Mittal show that this game can be won perfectly with a quantum strategy in the tensor product framework if and only if there exists a finite dimensional operator solution to the (multiplicative form of the) LBCS. An \emph{operator solution} is an assignment of selfadjoint operators $X_i$ on some Hilbert space $\mathcal{H}$ to the variables $x_i$ such that $X_i^2 = I$ for all $i \in [n]$, and such that all of the operators appearing in a constraint commute with each other (this removes any ambiguity in the order of the multiplication in the constraint). Furthermore, on the right-hand sides the $(-1)^{b_\ell}$ is replaced with $(-1)^{b_\ell} I$.  Such an operator solution is \emph{finite-dimensional} if $\mathcal{H}$ is finite-dimensional. The nonlocal game corresponding to the LBCS in~(\ref{eqn:BCS}) is the well known Mermin-Peres magic square game~\cite{mermin}, which can be won perfectly with a quantum strategy, and thus such a finite-dimensional operator solution does in fact exist.

The results of~\cite{clevemittal} were extended to the quantum commuting framework in~\cite{cleveslofstra}. Specifically, they showed that the game corresponding to an LBCS has a perfect quantum strategy in the commuting framework if and only if the LBCS has a possibly infinite dimensional operator solution.

In~\cite{qiso1}, the authors introduce a construction that takes any LBCS $\mathcal{F}$ and produces a graph $X(\mathcal{F})$. They show that, if $\mathcal{F}_0$ is the LBCS obtained from $\mathcal{F}$ by changing the right-hand side of every constraint in $\FF$ to 0 (known as the \emph{homogeneous} version of $\FF$), then the graphs $X(\FF)$ and $X(\FF_0)$ are quantum commuting isomorphic if and only if $\FF$ has an operator solution, and quantum tensor isomorphic if and only if $\FF$ has a finite-dimensional operator solution. Moreover, $X(\FF)$ and $X(\FF_0)$ are isomorphic if and only if $\FF$ is satisfiable. This allowed them to construct non-isomorphic graphs which are nevertheless quantum tensor/commuting isomorphic. This is (part of) the construction we will use to find quantum vertex transitive graphs which are not vertex transitive, thus we now give a description of this construction.

Let $\FF$ be an LBCS with constraints $C_\ell$ given by $\sum_{x_i \in S_\ell} x_i = b_\ell$ for $\ell \in [m]$. Then the graph $X(\FF)$ has as its vertices pairs $(\ell, f)$ where $\ell \in [m]$ and $f: S_\ell \to \mathbb{F}_2$ is an assignment of values to the variables in $C_\ell$ such that it is satisfied. There is an edge between two vertices $(\ell, f)$ and $(k,f')$ if they are \emph{inconsistent}, i.e., if there exists $x_i \in S_\ell \cap S_k$ such that $f(x_i) \ne f'(x_i)$. Note that this implies that the vertices of the form $(\ell,f)$ for fixed $\ell \in [m]$ form a clique in $X(\FF)$.

\begin{figure}[h!]
\begin{center}
\includegraphics[scale=.52]{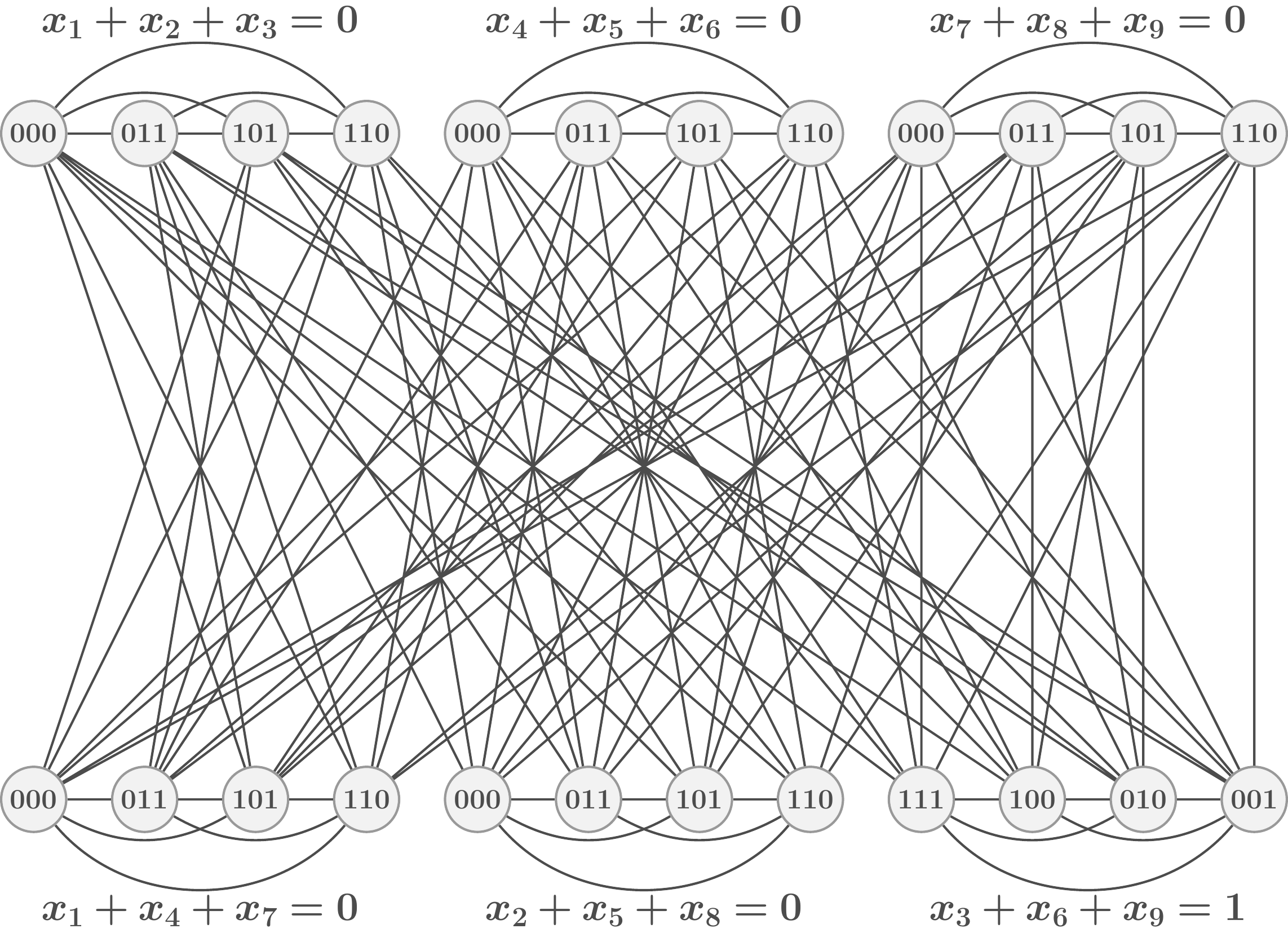}
\caption{$X(\FF)$ for the Mermin magic square game~(\ref{eqn:BCS}).}\label{fig:qiso1}
\end{center}
\end{figure}

\begin{figure}[h!]
\begin{center}
\includegraphics[scale=.52]{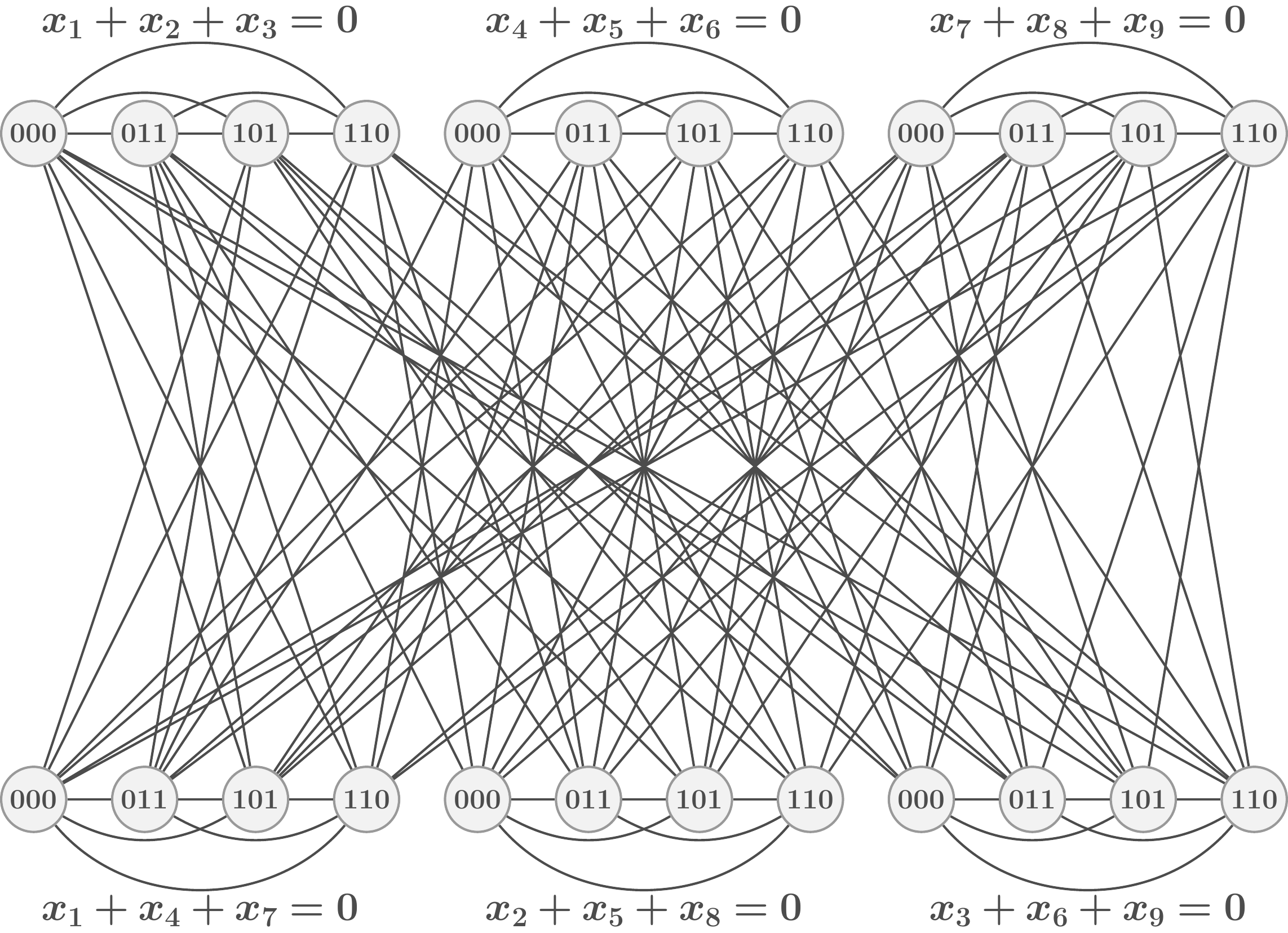}
\caption{$X(\FF_0)$ for the Mermin magic square game~(\ref{eqn:BCS}).}\label{fig:qiso2}
\end{center}
\end{figure}

Letting $\FF$ be the LBCS in~(\ref{eqn:BCS}), one obtains the two graphs from figures~\ref{fig:qiso1} and~\ref{fig:qiso2}, which were first considered in~\cite{qiso1}, and are quantum tensor isomorphic but not isomorphic. As was mentioned in~\cite{qiso1}, both of these graphs are vertex transitive, and even Cayley graphs. Moreover, if one removes the edges among the cliques corresponding to each constraint, the graphs remain quantum isomorphic but not isomorphic, and also become arc transitive (the automorphism groups act transitively on ordered pairs of adjacent vertices). In the next section we will introduce a construction of Arkhipov~\cite{arkhipov} which takes a connected non-planar graph and produces a LBCS with a (finite dimensional) operator solution but no classical solution. This construction is the other half of our construction of quantum vertex transitive graphs that are not vertex transitive.

\subsection{The construction of Arkhipov}\label{subsec:arkhipov}

In~\cite{arkhipov}, Arkhipov finds a correspondence between graphs and LBCS's in which every variable appears in exactly two equations. In the case of connected graphs, he shows that the corresponding LBCS is never satisfiable, but has an operator solution if and only if the graph is non-planar, in which case the operator solution is in finite dimension (even dimension at most eight). The construction is as follows: Let $Z$ be a connected graph with vertex set $[m]$ and label the edges $1, \ldots, n := |E(X)|$. Pick a vertex $\ell^* \in [m]$. Create an LBCS $\FF$ which will have a constraint $C_\ell$ for each $\ell \in [m]$ and variable $x_i$ for each edge $i \in [n]$. Let $S_\ell = \{x_i : \text{edge } i \text{ is incident to vertex } \ell\}$, and define $C_\ell$ to be the constraint $\sum_{x_i \in S_\ell} x_i = 0$ for all $\ell \ne \ell^*$, and define $C_{\ell^*}$ to be $\sum_{x_i \in S_{\ell}^*} x_i = 1$.

We remark that though the LBCS constructed depends on the choice of $\ell^*$, the satisfiability (classical or quantum) does not. In fact Arkhipov showed that the satisfiability only depends on the parity of the number of constraints whose right-hand side is 1. We will mainly be interested in the case where the graph $Z$ used to construct the LBCS is vertex transitive, so even the LBCS will not depend on the choice of $\ell^*$ (up to relabelling).

The two prototypical non-planar graphs are the complete graph $K_5$ and complete bipartite graph $K_{3,3}$. For $Z = K_{3,3}$, the corresponding LBCS is the Mermin-Peres magic square one from~(\ref{eqn:BCS}). For $Z = K_5$, the LBCS corresponds to the related magic pentagram game.

To construct quantum vertex transitive graphs which are not vertex transitive, we will concatenate Arkhipov's construction of an LBCS with the reduction from~\cite{qiso1} of an LBCS to graph isomorphism. It is thus helpful to give a direct description of the resulting two graphs based on the input graph to Arkhipov's construction.

\begin{definition}
Given a connected graph $Z$ with marked vertex $\ell^*$, we define two graphs $X(Z,\ell^*)$ and $X_0(Z)$ as follows. The vertices of $X(Z,\ell^*)$ are ordered pairs $(\ell,S)$ where $\ell \in V(Z)$ and $S$ is a subset of edges incident to $\ell$ which has even parity if $\ell \ne \ell^*$ and odd parity otherwise. Two vertices $(\ell,S)$ and $(k,T)$ are adjacent if $\ell$ and $k$ are adjacent and the edge $\ell k$ is in exactly one of $S$ and $T$. The graph $X_0(Z)$ is defined in the same way except that the subset $S$ must have even parity for all vertices $(\ell,S)$ of $X_0(Z)$ (thus it does not depend on $\ell^*$). In cases where the choice of $\ell^*$ is not of particular importance, such as when $Z$ is vertex transitive, we will sometimes simply write $X(Z)$.
\end{definition}

The following is merely the concatenation of results from~\cite{arkhipov} and~\cite{qiso1}:

\begin{theorem}\label{thm:concat}
Let $Z$ be a connected non-planar graph. Then the graphs $X_0(Z)$ and $X(Z,\ell^*)$ are quantum tensor isomorphic but not isomorphic for any choice of $\ell^* \in V(Z)$.
\end{theorem}

\subsection{Vertex transitivity of $X_0(Z)$}\label{subsec:vtxtrans}

In this section we will show that if $Z$ is vertex transitive, then $X_0(Z)$ is also vertex transitive. The first lemma shows that the automorphisms of $Z$ can be used to construct certain automorphisms of $X_0(Z)$. To state this lemma as simply as possible, we will use the convention that if $e = \ell k$ is an edge of $Z$ and $\sigma$ an automorphism of $Z$, then $\sigma(e)$ is defined as the edge $\sigma(\ell)\sigma(k)$. Moreover, if $S$ is a subset of edges, then $\sigma(S) := \{\sigma(e) : e \in S\}$.

\begin{lemma}\label{lem:Zauts}
Let $Z$ be a connected graph. If $\sigma \in \aut(Z)$, then the map $(\ell,S) \mapsto (\sigma(\ell), \sigma(S))$ is an automorphism of $X_0(Z)$.
\end{lemma}
\proof
Let us call the map described in the lemma statement $\hat{\sigma}$. First, we must show that this maps vertices of $X_0(Z)$ to vertices of $X_0(Z)$. However, this is clear since if $e$ is an edge incident to $\ell$, then $e = \ell k$ for some $k \in V(Z)$ and then $\sigma(e) = \sigma(\ell)\sigma(k)$ is incident to $\sigma(\ell)$. Thus $\sigma(S)$ is a subset of edges incident to $\sigma(\ell)$ whenever $S$ was a subset of edges incident to $\ell$. Moreover, $\sigma$ is a bijection on the vertices as well as the edges, and thus $|\sigma(S)| = |S|$. So $\sigma(S)$ has the same parity as $S$. This shows that the map $\hat{\sigma}$ does in fact map vertices of $X_0(Z)$ to vertices of $X_0(Z)$. Moreover, this map is a bijection, which is straightforward to see.

Next we must show that it preserves the adjacency of vertices in $X_0(Z)$. For this, suppose that $(\ell,S)$ and $(k,T)$ are adjacent. Then $e = \ell k$ is an edge of $Z$ and without loss of generality we have that $e \in S \setminus T$. From here we see that $\sigma(e) = \sigma(\ell) \sigma(k)$ is an edge of $Z$ and that $\sigma(e) \in \sigma(S) \setminus \sigma(T)$ since $\sigma$ is a bijection on the edges of $Z$. Therefore, $(\sigma(\ell),\sigma(S))$ is adjacent to $(\sigma(k), \sigma(T))$ in $X_0(Z)$. Since $\hat{\sigma}$ is a bijection that preserves adjacency, it also preserves non-adjacency and is therefore a bijection.\qeds

The above lemma shows that if there is an automorphism of $Z$ mapping $\ell$ to $k$, then there is an automorphism of $X_0(Z)$ which maps the subset of vertices with first coordinate $\ell$ to the subset of vertices with first coordinate $k$. The next lemma shows that we can use the even subgraphs of $Z$ to construct some automorphisms of $X_0(Z)$ which fix the first coordinate. An \emph{even subgraph} is a subgraph in which every vertex has even degree. We use the notation $S \triangle T$ to denote the symmetric difference of sets $S$ and $T$. We will also use $E(\ell)$ to denote the set of edges incident to the vertex $\ell$.

\begin{lemma}\label{lem:evensub}
Let $Z$ be a connected graph and $F$ the edge set of some even subgraph of $Z$. Then the map $(\ell,S) \mapsto (\ell, S \triangle (E(\ell) \cap F))$ is an automorphism of $X_0(Z)$.
\end{lemma}
\proof
Let $f$ denote the map describe in the lemma statement. The idea here is the even subgraphs of $Z$ correspond exactly to the solutions to the homogeneous version of the LBCS constructed from $Z$ via Arkhipov's method. The sum of two solutions to a homogeneous LBCS is a solution, and this addition corresponds to symmetric difference in subset language.

First we show that $f$ is indeed a map from the vertices of $X_0(Z)$ to the vertices of $X_0(Z)$. Since $F$ is the edge set of an even subgraph of $Z$, we have that $E(\ell) \cap F$ has even parity for all $\ell \in V(Z)$. Therefore, if $S$ is an even subset of $E(\ell)$, then $S \triangle (E(\ell) \cap F)$ must have even parity since the parity of the symmetric difference is the sum of the parities. Thus $f$ is a map from $V(X_0(Z))$ to itself, and then it is easy to see that it must be a bijection, since taking symmetric difference with a fixed set is a bijection on subsets. In fact, $f$ is a bijection on the vertices with fixed first coordinate.

Next we must show that $f$ preserves adjacency. For this suppose that $(\ell, S)$ and $(k,T)$ are adjacent vertices of $X_0(Z)$. As in the proof of the previous lemma we have that $e = \ell k$ is an edge of $Z$ and without loss of generality we have that $e \in S \setminus T$. Note that $e \in E(\ell)$ and $e \in E(k)$. If $e \not\in F$, then $e \notin E(\ell) \cap F$ and thus $e \in S \triangle (E(\ell) \cap F)$ but $e \not\in T \triangle (E(k) \cap F)$. If $e \in F$, then $e \in E(\ell) \cap F$ and $e \in E(k) \cap F$, thus $e \not\in S \triangle (E(\ell) \cap F)$ but $e \in T \triangle (E(k) \cap F)$. In either case we see that $(\ell, S \triangle (E(\ell) \cap F))$ is adjacent to $(k, T \triangle (E(k) \cap F))$, and so $f$ preserves adjacency and we are done.\qeds

In the case where $Z$ has no cut-vertex, we can use the above to show that any vertex $(\ell,S)$ of $X_0(Z)$ can be mapped to a vertex of the form $(\ell, T)$ by an automorphism of $X_0(Z)$. A cut-vertex of a connected graph is a vertex whose removal disconnects the graph.

\begin{lemma}\label{lem:cutvtx}
Let $Z$ be a connected graph with no cut-vertex. If $(\ell,S)$ and $(\ell,T)$ are two vertices of $X_0(Z)$, then there is an automorphism $\sigma$ of $X_0(Z)$ such that $\sigma(\ell,S) = (\ell,T)$.
\end{lemma}
\proof
First we will show that the lemma holds when $|S \triangle T| = 2$. One can then compose such automorphisms to prove the result for any $|S \triangle T|$.

Suppose that $S \triangle T = \{e,f\}$ for edges $e,f$ incident to $\ell$, and let $j,k \in V(Z)$ be such that $e = \ell j$ and $f = \ell k$. Consider the graph $Z'$ obtained from $Z$ by removing vertex $\ell$. Since $\ell$ is not a cut-vertex by assumption, the graph $Z'$ is connected, and therefore there exists a path $P$ from $j$ to $k$ in $Z'$. This path does not contain any edge incident to $\ell$ since these edges are not in $Z'$. In $Z$, the path $P$ along with edges $e,f$ and vertex $\ell$ form a cycle $C$. Let $F$ be the edge set of this cycle and note that $E(\ell) \cap F = \{e,f\}$. Since $C$ is a cycle, it is an even subgraph of $Z$. Therefore, by Lemma~\ref{lem:evensub}, the map $(k,R) \mapsto (k,R \triangle (E(k) \cap F))$ is an automorphism of $Z$. It is easy to see that $S \triangle (E(\ell) \cap F) = S \triangle \{e,f\} = T$, and therefore this automorphism maps $(\ell, S)$ to $(\ell,T)$ as desired.

For $|S \triangle T| > 2$, we will still have that this cardinality is even (since $S$ and $T$ are even), and so we can partition $S \triangle T$ into pairs $\{e_i,f_i\}$ for $i = 1, \ldots, |S \triangle T|/2$. Defining $S_0 = S$ and letting $S_i = S_{i-1} \triangle \{e_i,f_i\}$, it is easy to see that $|S_{i-1}\triangle S_i| = |\{e_i,f_i\}| = 2$ for all $i$, and $S_{|S \triangle T|/2} = T$. Therefore we can compose automorphisms mapping $(\ell,S_{i-1})$ to $(\ell,S_i)$ guaranteed by the above argument to obtain an automorphism mapping $(\ell,S)$ to $(\ell,T)$.\qeds

We actually only needed to assume that $\ell$ was not a cut-vertex in the above lemma, but the statement as is suffices for our results.

So Lemma~\ref{lem:Zauts} provides us with automorphisms of $X_0(Z)$ that change the first coordinate (and possibly the second coordinate), and Lemma~\ref{lem:evensub} provides us with automorphisms that fix the first coordinate but change the second coordinate. Combining these we can prove the following:

\begin{theorem}\label{thm:XZvtxtrans}
Let $Z$ be a connected vertex transitive graph. Then $X_0(Z)$ is vertex transitive.
\end{theorem}
\proof
We must prove that $Z$ has no cut-vertex and then it will follow easily from the above lemmas. Since $Z$ is connected, it has a spanning tree $\mathcal{T}$. Consider a leaf (vertex of degree 1) $\ell$ of $\mathcal{T}$, which must exist. Removing $\ell$ from $\mathcal{T}$ creates a smaller tree $\mathcal{T}'$ which must be connected since it is a tree. Since $\mathcal{T}'$ is a subgraph of the graph $Z'$ obtained by removing $\ell$ from $Z$, we have that $Z'$ must be connected as well. Therefore, $\ell$ was not a cut-vertex. However, $Z$ is vertex transitive, so if $\ell$ is not a cut-vertex, then no vertex of $Z$ is a cut-vertex.

Now consider vertices $(\ell,S)$ and $(k,T)$ of $X_0(Z)$. Since $Z$ is vertex transitive, there exists an automorphism $\sigma \in \aut(Z)$ such that $\sigma(\ell) = k$. By Lemma~\ref{lem:Zauts}, there is an automorphism $\sigma_1$ of $X_0(Z)$ such that $\sigma_1(\ell,S) = (k,T')$ where $T' = \sigma(S)$. Now, since $Z$ has no cut-vertex, Lemma~\ref{lem:cutvtx} says that there is an automorphism $\sigma_2$ of $X_0(Z)$ such that $\sigma_2(k,T') = (k,T)$. Therefore, the composition $\sigma_2 \circ \sigma_1$ is an automorphism of $X_0(Z)$ that maps $(\ell,S)$ to $(k,T)$. Since these vertices were arbitrary, this implies that $X_0(Z)$ is vertex transitive as desired.\qeds

\begin{corollary}
If $Z$ is a connected vertex transitive non-planar graph, then $X(Z, \ell^*)$ is quantum vertex transitive for any choice of $\ell^*$.
\end{corollary}
\proof
Under these conditions, the graphs $X_0(Z)$ and $X(Z)$ are quantum isomorphic by Theorem~\ref{thm:concat}. This implies that they have the same number of quantum orbits. However, $X_0(Z)$ is vertex transitive by Theorem~\ref{thm:XZvtxtrans} and so it has only one orbit and therefore only one quantum orbit. Thus $X(Z)$ has one quantum orbit, i.e., is quantum vertex transitive.\qeds

It would be interesting to find a connected non-planar graph $Z$ such that exactly one of $X_0(Z)$ and $X(Z)$ was vertex transitive.

\subsection{Quantum vertex transitivity of the disjoint union}\label{subsec:disjoint}

In this section we aim to show that if $Z$ is a connect non-planar vertex transitive graph, then the disjoint union of $X_0(Z)$ and $X(Z)$ is quantum vertex transitive but not vertex transitive. To begin we need the following lemma:

\begin{lemma}\label{lem:qiso2qtrans}
Let $X$ and $Y$ be quantum isomorphic graphs which are both quantum vertex transitive. Then the disjoint union of $X$ and $Y$ is quantum vertex transitive.
\end{lemma}
\proof
We denote by $A_X$ and $A_Y$ the adjacency matrices of $X$ and $Y$, respectively. Let $Z$ denote the disjoint union of $X$ and $Y$, and let $U = (u_{xx'})$ and $V = (v_{yy'})$ be the magic unitaries whose entries are the canonical generators of $\A(X)$ and $\A(Y)$ respectively. Define $W = (w_{zz'}) = U \oplus V$ as
\[w_{zz'} = \begin{cases}
u_{zz'} \otimes 1_{\A(Y)} , & \text{if } z,z' \in V(X) \\
1_{\A(X)} \otimes v_{zz'} , & \text{if } z,z' \in V(Y) \\
0, & \text{otherwise.}
\end{cases}
\]
It is easy to see that this is a magic unitary. We aim to show that it commutes with the adjacency matrix of the disjoint union of $X$ and $Y$, which we will denote by $A$. Suppose that $z_1,z_2,z'_1,z'_2 \in V(Z)$ and $\rel(z_1,z_2) \ne \rel(z'_1,z'_2)$. We aim to show that $w_{z_1z'_1}w_{z_2z'_2} = 0$. If all four vertices are contained in one of $V(X)$ or $V(Y)$, then it is easy to see that $w_{z_1z'_1}w_{z_2z'_2} = 0$ by the properties of the magic unitaries $U$ and $V$. Therefore we may assume without loss of generality that $z_1 \in V(X)$. If $z'_1 \not\in V(X)$, then $u_{z_1z'_1} = 0$ by definition and we are done. So we can assume that $z'_1 \in V(X)$. Since not all four vertices are contained in $V(X)$ we must have that $z_2$ and $z'_2$ are contained in $V(Y)$. However, this implies that $z_1$ and $z_2$ are distinct nonadjacent vertices, and so are $z_2$ and $z'_2$. This contradicts our assumption and so we are done. This implies that $W$ does in fact commute with $A$. Therefore, if $w_{zz'} \ne 0$, then $z \sim_1 z'$, i.e., $z$ and $z'$ are in the same quantum orbit of $Z$. By assumption we have that both $X$ and $Y$ are quantum vertex transitive, i.e., have one quantum orbit. Therefore, $u_{xx'} \ne 0$ for all $x,x' \in V(X)$ and $v_{yy'} \ne 0$ for all $y,y' \in V(Y)$. Therefore $w_{xx'} \ne 0$ for all $x,x' \in V(X)$ and $w_{yy'} \ne 0$ for all $y,y' \in V(Y)$, i.e., every vertex of $X$ is in the same quantum orbit of $Z$ and every vertex of $Y$ is in the same quantum orbit of $Z$.

It now only remains to show that $x \sim_1 y$ for some $x \in V(X)$ and $y \in V(Y)$. Since we assumed that $X$ and $Y$ are quantum isomorphic, there exists a unital C*-algebra $\A$ and magic unitary $\hat{W} = (\hat{w}_{xy})$ for $x \in V(X)$, $y \in V(Y)$ such that $A_X\hat{W} = \hat{W}A_Y$, i.e., $\hat{w}_{xy}\hat{w}_{x'y'} = 0$ whenever $\rel(x,x') \ne \rel(y,y')$. Define $\tilde{W} = (\tilde{w}_{zz'})$ for $z,z' \in V(Z)$ as
\[\tilde{w}_{zz'} = \begin{cases}
\hat{w}_{zz'} , & \text{if } z \in V(X), z' \in V(Y) \\
\hat{w}_{z'z} , & \text{if } z \in V(Y), z' \in V(X) \\
0, & \text{otherwise.}
\end{cases}
\]
It is easy to see that $\tilde{W}$ is a magic unitary. Suppose that $z_1,z_2,z'_1,z'_2 \in V(Z)$ and $\rel(z_1,z_2) \ne \rel(z'_1,z'_2)$. We aim to show that $\tilde{w}_{z_1z'_1}\tilde{w}_{z_2z'_2} = 0$. If $z_1$ and $z'_1$ are both contained in $X$ or both contained in $Y$, then $\tilde{w}_{z_1z'_1} = 0$ and we are done. So we may assume without loss that $z_1 \in V(X)$ and $z'_1 \in V(Y)$, so let us rename these variable $x$ and $y$ respectively. Similarly, we either have $z_2 \in V(X)$ and $z'_2 \in V(Y)$, or $z_2 \in V(Y)$ and $z'_2 \in V(X)$. In either case, let us rename the variable as $x'$ and $y'$ according to which graph they are contained in. In the first of the two cases, this results in $\rel(x,x') \ne \rel(y,y')$ and we have $\tilde{w}_{xy}\tilde{w}_{x'y'} = \hat{w}_{xy}\hat{w}_{x'y'} = 0$ since $\hat{W}$ was a magic unitary witnessing a quantum isomorphism of $X$ and $Y$. In the second case, we have $\rel(x,y) \ne \rel(x'y')$, but this is a contradiction since every vertex of $X$ is distinct and nonadjacent to every vertex of $Y$ in their disjoint union $Z$. Therefore, $\tilde{W}$ is a magic unitary which commutes with $A$. Moreover, by its definition, for any vertex $x \in V(X)$, there must be at least one vertex $y \in V(Y)$ such that $\tilde{w}_{xy} \ne 0$. Thus these two vertices are in the same quantum orbit, and combining this with the above we see that there is only one quantum orbit of $Z$, i.e, it is quantum vertex transitive.\qeds

We can now prove our main result of this section:

\begin{theorem}
Let $X$ and $Y$ be quantum vertex transitive graphs which are quantum isomorphic but not isomorphic. Then the disjoint union of $X$ and $Y$ are quantum vertex transitive but not vertex transitive.
\end{theorem}
\proof
By Lemma~\ref{lem:qiso2qtrans}, we have that the disjoint union of $X$ and $Y$ is quantum vertex transitive. So it is only left to show that it is not vertex transitive. It is easy to see that a graph is vertex transitive if and only if all of its connected components are vertex transitive and isomorphic to each other (since automorphism must map connected components to connected components). Thus, if the disjoint union of $X$ and $Y$ is vertex transitive, then $X$ and $Y$ must have both been the disjoint union of (a possibly different) number of copies of a common connected vertex transitive graph $Z$. However, since we assumed that $X$ and $Y$ were not isomorphic, this implies that the number of copies of $Z$ used was different for $X$ and $Y$. This implies that $X$ and $Y$ have a different number of vertices, and therefore cannot be quantum isomorphic, a contradiction to our assumption. Therefore the disjoint union of $X$ and $Y$ cannot be vertex transitive and we are done.\qeds

As a corollary we obtain the following which allows us to construct an infinite number of graphs which are quantum vertex transitive but not vertex transitive.

\begin{corollary}\label{cor:construct}
Let $Z$ be a connected, vertex transitive, non-planar graph. Then the disjoint union of $X_0(Z)$ and $X(Z)$ is quantum vertex transitive but not vertex transitive.
\end{corollary}

The smallest possibility for $Z$ in the above corollary is the graph $K_5$, but the resulting graphs $X_0(Z)$ and $X(Z)$ are actually larger (40 vertices) than those obtained from letting $Z = K_{3,3}$ which have 24 vertices and are shown in Figures~\ref{fig:qiso1} and~\ref{fig:qiso2}. By Kuratowski's theorem, any other non-planar graph must contain one of $K_{3,3}$ or $K_5$ as a minor and so these are indeed the smallest graphs that can be obtained by this construction. As we mentioned previously, if one removes the edges contained in the cliques corresponding to the constraints of the LBCS, then the graphs in Figures~\ref{fig:qiso1} and~\ref{fig:qiso2} remain quantum isomorphic but not isomorphic, and now become arc-transitive. Therefore, their disjoint union will be quantum arc-transitive, but not arc-transitive, or even edge-transitive.

There are infinitely many connected, vertex transitive, non-planar graphs and so the above corollary does in fact give us infinitely many examples of graphs which are quantum vertex transitive but not vertex transitive. Indeed, any vertex transitive graph with degree at least six is non-planar since it is known that any planar graph has some vertex of degree less than six.

We remark that if one would prefer an example of a \emph{connected} graph which is quantum vertex transitive but not vertex transitive, then they may simply take the complement of the examples generated by Corollary~\ref{cor:construct}.

As we mentioned, the above corollary answers an unresolved question raised (at least implicitly) in the quantum permutation group literature (see e.g.~\cite{banicahomogeneous, banica05small}). Perhaps the examples provided by this corollary will have some new interesting quantum permutation groups. Unfortunately it seems that the currently known techniques to compute quantum isomorphism groups are not sufficient to handle graphs of this size. Hopefully this will change with the increased interest in this area from members of the quantum information and graph theory community, which we hope this work encourages.

\section{Discussion}\label{sec:discuss}

Here we discuss some possible further directions and open questions.

\paragraph{No quantum symmetry.}

In the study of quantum automorphism groups of graphs, it seems an important case is when the graph has no quantum symmetry, i.e., its quantum symmetry algebra is commutative. Indeed, many of the small graphs whose quantum automorphism groups have been computed~\cite{qauts11,noqsymm} have this property. The most direct way to show this is to show that all of the entries $u_{ij}$ of the magic unitary $U$ defining the given quantum automorphism group commute. However, in the known examples a different strategy is followed. The most common technique seems to be to use the universal action.

Suppose that $\mathcal{Q}$ is a quantum permutation group on the set $[n]$ with universal action $u$. It was shown in~\cite{banicahomogeneous} that the magic unitary $U$ commutes with the projection onto a subspace $K$ of $\mathbb{C}^n$ if and only if $K$ is invariant for $u$, i.e.\ $u(K) \subseteq K \otimes C(\mathcal{Q})$. If $\mathcal{Q} = \qut(X)$ for some graph $X$ with adjacency matrix $A$, then the corresponding magic unitary $U$ commutes with $A$ and therefore any polynomial expression in $A$. This includes the projections onto the eigenspaces of $A$. Thus if $f_1, \ldots, f_d$ is a basis of one of these eigenspaces, then $u(f_j) = \sum_{i=1}^d f_i \otimes a_{ij}$ for some elements $a_{ij} \in C(\mathcal{Q})$. Since the image of $u$ is dense in $\mathcal{Q}$, if all of the $a_{ij}$ (as well as those for the other eigenspaces) commute, then $\mathcal{Q}$ is commutative and therefore $X$ has no quantum symmetry. Our results show that one does not necessarily need to consider the eigenspaces of the adjacency matrix, but instead could use any Hermitian matrix in the coherent algebra of the graph. This will often allow for smaller eigenspaces than considering the adjacency matrix alone.

The above method has been particularly useful for \emph{circulants}: graphs $X$ with vertex set $\mathbb{Z}_n$ such that $i \sim j$ if $i - j \in C$ for some inverse closed subset $C \subseteq \mathbb{Z}_n \setminus \{0\}$. The reason this approach works so well for these graphs is because they always have a basis of eigenvectors that is closed under entrywise product. Specifically, if $f \in \mathbb{C}^n$ is the vector whose $i^\text{th}$ entry is $\omega^i$ for a primitive $n^\text{th}$ root of unity $\omega$, then the entrywise powers $f^j$ for $j = 0, \ldots, n-1$ are an orthonormal basis of $\mathbb{C}^n$ consisting of eigenvectors of $X$. It is known that the vectors $f^j$ and $f^{n-j}$ always correspond to the same eigenvalue for these graphs. It was shown in~\cite{banicahomogeneous} that if these are the only common eigenvalues among the vectors $f^j$ for $j = 0, \ldots, n-1$ then the graph $X$ has no quantum symmetry (unless $n = 4)$. Here again, our results could be of use. In the case of circulants, the resulting coherent algebra only contains symmetric matrices, and therefore is a \emph{symmetric association scheme}. This means that all of its elements commute and therefore have a common eigenbasis. In fact the common eigenbasis is exactly the one described above consisting of the $f^j$. It is known that there will be a matrix in the coherent algebra with different eigenvalues for $f^j$ and $f^i$ whenever $i \not\in\{j,n-j\}$ if and only if there are $\lfloor n/2 \rfloor + 1$ classes in the coherent algebra. In this case the classes are exactly the distance relations of the cycle of length $n$. Thus, if the coherent configuration of a circulant has $\lfloor n/2 \rfloor + 1$ classes then the circulant has no quantum symmetry.

Unfortunately, using coherent algebras does not provide any further information for a certain class of graphs: strongly regular graphs. The coherent configurations of these graphs have only three classes: the vertices, edges, and non-edges. Thus the coherent algebra is simply the span of the identity $I$, the adjacency matrix $A$, and the adjacency matrix of the complement $\overline{A} = J-I-A$. This means that Corollary~\ref{cor:extrarels} is just a restatement of the relations defining the quantum automorphism group. It may be possible that the quantum orbitals are finer than this, but we do not know how to compute these in general. Moreover, it is possible that the classes of the coherent configuration and the (classical) orbitals of the graph coincide, and therefore so do the quantum orbitals. 

\paragraph{Higher order orbits.} In this work we defined the orbits of a quantum permutation group on points and ordered pairs of points. One could also define a relation $\sim_m$ on ordered $m$-tuples of points such that $(x_1, \ldots, x_m) \sim_m (y_1, \ldots, y_m)$ whenever $u_{x_1y_1}\ldots u_{x_m y_m} = 0$, and then used this to define orbits of $m$-tuples for quantum permutation groups. We could have defined such a relation, but even for $m = 3$ we do not see how to show that it would be transitive (and do not believe it will be in general). It seems that things work for $m \le 2$ because we can multiply on both the left and the right, but for $m \ge 3$ we cannot multiply ``in the middle" since we do not have commutativity. But what about the equivalent definitions of orbits and orbitals given in Section~\ref{subsec:equivdef} using the canonical action? Is it possible to extend this in order to define orbits on $m$-tuples for $m \ge 3$?

Let $U = (u_{ij})$ be a magic unitary whose entries are the canonical generators of $C(\mathcal{Q})$ from some quantum permutation group $\mathcal{Q}$. The $m$-th tensor power of the action $u$ of $\mathcal{Q}$ is a linear map $u^{\otimes m} : (\mathbb{C}^n)^{\otimes m} \to(\mathbb{C}^n)^{\otimes m} \otimes C(\mathcal{Q})$ defined on the standard basis vectors as
\[
u^{\otimes m}(e_{i_1} \otimes \ldots \otimes e_{i_m}) = \sum_{j_1, \ldots, j_m} (e_{j_1} \otimes \ldots \otimes e_{j_m}) \otimes u_{j_1i_1} \ldots u_{j_m i_m}.\]
As with $u$ and $u^{\otimes 2}$, we can consider the space of fixed points of $u^{\otimes m}$:
\[
\mathrm{Hom}(1,u^{\otimes m})=\{f \in (\mathbb{C}^n)^{\otimes m} : u^{\otimes m}(f) = f \otimes 1\}.\]
In the case of $m = 2$, we saw that $\mathrm{Hom}(1,u^{\otimes 2})$ is a selfadjoint subalgebra of $\mathbb{C}^{X}\otimes \mathbb{C}^{X}$, and the same holds for $m=1$ since $u$ is an action. Moreover, we could define the orbits as the parts of the partition corresponding to $\mathrm{Hom}(1,u)$, and the orbitals as the parts of the partition corresponding to $\mathrm{Hom}(1,u^{\otimes 2})$. Unfortunately, we do not know whether $\mathrm{Hom}(1,u^{\otimes m})$ is a selfadjoint subalgebra of $(\mathbb{C}^{X})^{\otimes m}$. Thus we leave it as an open question how to define orbits of a quantum permutation group on $m$-tuples for general $m$.

\paragraph{Quantum groups vs.~quantum behaviour.}

Let $Z$ be the graphs consisting of a single edge plus two additional isolated vertices. In~\cite{qsymscalgs} it is shown (and it is not hard to see) that the quantum symmetry algebra of $Z$ is generated by a magic unitary of the form
\[
\begin{pmatrix}
a & 1 - a & 0 & 0 \\
1 - a & a & 0 & 0 \\
0 & 0 & b & 1 - b \\
0 & 0 & 1 - b & b
\end{pmatrix}
\]
where $a$ and $b$ are free projectors. Obviously, this is not commutative since $a$ and $b$ do not commute in general. Thus the graph $Z$ has some ``non-classical" automorphisms. Does this mean that we can construct a perfect quantum strategy for the $(Z,Z)$-isomorphism game (henceforth referred to as the $Z$-automorphism game) which is not classical? It depends on what exactly it means for a strategy to not be classical. For any choice of finite dimensional projectors $a$ and $b$, we can construct a perfect strategy for the $Z$-automorphism game in which the measurements used by the players will correspond to the rows and columns of the magic unitary above, and the shared state will be the maximally entangled state. This is of course not a classical strategy since it uses quantum measurements, but this does not mean that the corresponding behaviour of the players, i.e., the correlations of their inputs and outputs in the game, cannot be explained classically.

To be more precise, any fixed strategy (classical, quantum, or otherwise) for a nonlocal game has a corresponding \emph{correlation} or joint conditional probability distribution. This is a function $p$ mapping tuples of Alice and Bob's inputs and outputs to $[0,1]$, where $p(y,y'|x,x')$ is the probability that Alice and Bob reply with $y$ and $y'$ respectively, upon receiving $x$ and $x'$. In the deterministic case, this is either 0 or 1, but access to shared randomness allows classical players to construct strategies whose correlations are any convex combination of deterministic ones. In the quantum commuting framework, the probability $p(y,y'|x,x')$ is given by the value of a tracial state $s$ on the product of the the projector corresponding to Alice outputting $y$ upon receiving $x$ and the projector corresponding to Bob outputting $x'$ upon receiving $y'$. In the case of the $Z$-automorphism game, this means that all of the correlation probabilities are determined by the value of $s(ab) \in [0,1]$. We do not present the proof here, but we have shown that no matter what value is chosen for this, the resulting correlation can be obtained by a classical strategy. Therefore, there is no perfect quantum strategy for the $Z$-automorphism game which actually exhibits non-classical behaviour in this sense.

One could therefore consider the problem of which graphs not only have quantum symmetries, but such quantum symmetries that give rise to strategies for the automorphism game of the graph whose corresponding correlations can \emph{not} be obtained via a classical strategy. Such a quantum strategy is said to exhibit \emph{nonlocality}, since its behavior cannot be explained by local hidden variable model. There are three different degrees of nonlocality that can be exhibited by a quantum correlation $p$: nonlocality, logical nonlocality, and pseudotelepathy/strong nonlocality. The first is as we defined it above: there is no classical strategy that produces $p$. Logical nonlocality occurs when there is no classical correlation $q$ such that $q(y,y'|x,x') = 0$ if and only if $p(y,y'|x,x') = 0$. Strong nonlocality is defined as there being no classical correlation $q$ such that $q(y,y'|x,x') = 0$ whenever $p(y,y'|x,x') = 0$, even allowing for $q$ to be zero in places where $p$ is not. For any quantum correlation $p$ exhibiting strong nonlocality, one can construct a nonlocal game which is won perfectly by $p$ but by no classical strategy: define the game so that the players lose on input/output tuples where $p$ is 0. This ability to perfectly win a game which has no perfect classical strategy is what motivates the term pseudotelepathy, since it appears as if the players are telepathic.

In the case of the isomorphism game, pseudotelepathy occurs exactly when there is a pair of non-isomorphic graphs $X,Y$ that are quantum tensor/commuting isomorphic. This means that the $X$-automorphism game is never a pseudotelepathy game, since every graph has the trivial automorphism. However, it is possible for an automorphism game to exhibit nonlocality. Computations by the third author and Robert \v{S}\'{a}mal show that already the complete graph on five vertices (and thus any larger number of vertices) admits quantum tensor automorphisms whose correlations are not classical. We do not know whether the complete graph on four vertices has this property.

It is also possible for an automorphism game to exhibit logical nonlocality. This happens exactly when there are pairs of vertices $(x,x')$ and $(y,y')$ that are in different (classical) orbitals of $X$, but there is a quantum correlation $p$ which wins the automorphism game perfectly and $p(y,y'|x,x') > 0$. Since the Haar state is tracial, the correlation $p(y,y'|x,x') = h(u_{xy}u_{x'y'})$ is quantum commuting correlation which wins the $X$-automorphism game perfectly. Moreover, by Theorem~\ref{thm:haar}, we have that $h(u_{xy}u_{x'y'}) \ne 0$ if and only if $u_{xy}u_{x'y'} \ne 0$ if and only if $(x,x')$ and $(y,y')$ are in the same orbital of $\qut(X)$. Thus the correlation arising from the Haar state for $\qut(X)$ exhibits logical nonlocality if and only if the orbitals and quantum orbitals of $X$ differ.
Examples of such graphs can be constructed by taking the disjoint union of any pair of non-isomorphic but quantum isomorphic graphs. 

\subsection*{Acknowledgements}
We would like to thank Alexandru Chirvasitu and Teodor Banica for pointing out to us that the Haar state of a quantum permutation group is tracial.

Part of this work was completed during the Focused Research Meeting at Queen's University Belfast in October 2017, which was funded by the Heilbronn Institute for Mathematical Research. ML is partially supported by the National Science Foundation Grant DMS-1600186. DR was supported by European Research Council Advanced Grant GRACOL, project no.~320812.

\bibliographystyle{plainurl}

\bibliography{Quantum Automorphisms.bbl}

\end{document}